\documentclass[fleqn,10pt]{wlscirep}
\usepackage[utf8]{inputenc}
\usepackage[T1]{fontenc}
\usepackage{graphicx}%
\usepackage{multirow}%
\usepackage{amsmath,amssymb,amsfonts}%
\usepackage{amsthm}%
\usepackage{mathrsfs}%
\usepackage{xcolor}%
\usepackage{textcomp}%
\usepackage{manyfoot}%
\usepackage{booktabs}%
\usepackage{algorithm}%
\usepackage{algorithmicx}%
\usepackage{algpseudocode}%
\usepackage{listings}%
\usepackage{subcaption}
\usepackage{algorithm}
\usepackage{algpseudocode}
\usepackage{listings}%
\usepackage{subcaption}
\usepackage{makecell}  
\usepackage{array}     

\newcommand{\argmin}{\mathop{\mathrm{argmin}}}

\title{Do we actually understand the impact of renewables on electricity prices? A causal inference approach}

\author[1,*]{Davide Cacciarelli}

\author[1,2,3,4]{Pierre Pinson}

\author[5]{Filip Panagiotopoulos}

\author[5]{David Dixon}

\author[5]{Lizzie Blaxland}

\affil[1]{Imperial College London, London, UK}

\affil[2]{Technical University of Denmark, Kgs. Lyngby, Denmark}

\affil[3]{Halfspace, Copenhagen,Denmark}

\affil[4]{Aarhus University, Aarhus, Denmark}

\affil[5]{National Energy System Operator, Wokingham, UK}

\affil[*]{d.cacciarelli@imperial.ac.uk}

\begin{abstract}
The energy transition is profoundly reshaping electricity market dynamics. It makes it essential to understand how renewable energy generation actually impact electricity prices, among all other market drivers. These insights are critical to design policies and market interventions that ensure affordable, reliable, and sustainable energy systems. However, identifying causal effects from observational data is a major challenge, requiring innovative causal inference approaches that go beyond conventional regression analysis only. We build upon the state of the art by developing and applying a local partially linear double machine learning approach. Its application yields the first robust causal evidence on the distinct and non-linear effects of wind and solar power generation on UK wholesale electricity prices, revealing key insights that have eluded previous analyses. We find that, over 2018-2024, wind power generation has a U-shaped effect on prices: at low penetration levels, a 1 GWh increase in energy generation reduces prices by up to 7 GBP/MWh, but this effect gets close to none at mid-penetration levels (20–30\%) before intensifying again. Solar power places substantial downward pressure on prices at very low penetration levels (up to 9 GBP/MWh per 1 GWh increase in energy generation), though its impact weakens quite rapidly. We also uncover a critical trend where the price-reducing effects of both wind and solar power have become more pronounced over time (from 2018 to 2024), highlighting their growing influence on electricity markets amid rising penetration. Our study provides both novel analysis approaches and actionable insights to guide policy makers in appraising the way renewables impact electricity markets.
\end{abstract}
\begin{document}

\flushbottom
\maketitle
%
%
\thispagestyle{empty}

\noindent
Increasing the share of electricity from renewable energy sources, e.g., wind energy and solar energy, is one of the key actions towards climate change mitigation. Projections from the International Energy Agency (IEA) indicate that by 2030, wind and solar power generation assets will jointly contribute to 40\% of global electric power generation \cite{IEA2023}. In the UK, wind energy alone represented 29. 4\% of total electricity generation in 2023 \cite{nationalgrid2023_wind}. Solar energy, while currently in a less prevalent position, is increasing its market share rapidly, driven by low technology costs and generous support policies. The integration of these volatile energy sources (also with limited predictability) is profoundly affecting electricity markets, requiring a deeper understanding of their impact on market dynamics. Although the environmental benefits of renewable energy generation, such as reduced CO\textsubscript{2} emissions, are well-documented \cite{IPCC2023Summary}, their increasing penetration introduces both opportunities and challenges for the operation of the electricity market \cite{morales2013integrating}. From a market perspective, the inherent lower marginal costs associated with wind and solar power generation are expected to exert downward pressure on wholesale day-ahead prices (also referred to as spot prices) by shifting the equilibrium point through the merit-order effect \cite{antweiler2021long}. However, from an operational point of view, the variability and limited predictability of renewable energy generation induce an additional need for system balancing and ancillary services \cite{miettinen2019impacts, goodarzi2019impact}. The market impact of renewable energy sources, particularly wind energy, has been the subject of extensive research over the past two decades. Early studies were conducted for the Danish electricity market (part of Nord sPool). Although actual wind power generation was found to have limited impact on prices \cite{morthorst2003wind}, forecast wind power penetration emerged as the primary determinant, reflecting the information available to market participants at the time of gate closure \cite{jonsson2010market}. Since these pioneering studies, researchers have conducted similar investigations in numerous other electricity markets, such as in Australia \cite{cutler2011high, forrest2013assessing, bell2017revitalising}, Germany \cite{wurzburg2013renewable, ketterer2014impact, gurtler2018effect, do2019impact, keeley2021impact}, Ireland \cite{o2011merit, o2014quantitative}, Italy \cite{clo2015merit}, Korea \cite{shcherbakova2014effect}, the Netherlands \cite{nieuwenhout2011impact}, the Nordics \cite{hu2021effects, spodniak2021impact, tselika2024quantifying}, Spain \cite{gelabert2011ex, cruz2011effect, azofra2014wind}the UK \cite{barthelmie2008economic, joos2018short}, and the US \cite{woo2011impact, woo2016merit, martinez2016impact, prol2020cannibalization}. More recently, the \emph{cannibalisation} effect, in which the market value of renewable energy declines as their penetration levels increase by displacing more expensive forms of generation, has been investigated for several European markets \cite{stiewe2024cross}. These studies highlight common trends observed across electricity markets globally, such as the price reduction resulting from the merit-order effect and the increased price volatility linked to higher levels of installed wind capacity. 

Despite these relevant insights, most of the aforementioned works rely on traditional regression-based approaches, which are often limited in their ability to fully capture the complexity of the impact of renewable energy on electricity markets. Here though, the diagnostic analytics approaches to be developed and the insights to be derived ought to be causal. Specifically, simpler analyses might struggle to unravel the influence of confounding factors, such as demand fluctuations, gas prices, and hourly price profiles. In addition, they often overlook the non-linear dynamics that govern the impact of renewable energy penetration on electricity prices, leading to biased or incomplete conclusions. Indeed, traditional econometric approaches are often constrained by their fully parametric nature, requiring strong assumptions about the functional form of the relationships being modelled. As a result, traditional methods often provide associations, but lack the causal clarity needed to inform policy making effectively. If not revealing true causal effects, consequent market interventions and new policies will be suboptimal at best, and completely inefficient in the worst case.

In contrast, our approach represents a significant methodological and analytical advancement. We introduce a bespoke causal inference framework that generalises double machine learning (DML) \cite{chernozhukov2018double} for use in a non-linear and non-stationary context. Our \emph{local partially linear DML} method enables the identification of non-linear treatment effects while accounting for a wide range of confounding factors. This allows us (i) to isolate the effects of predicted wind and solar power penetration from other influences, (ii) to examine how these impacts vary with predicted penetration levels, and (iii) to track their evolution over time. Focusing on the UK electricity market, which has seen a substantial rise in renewable energy penetration in recent years, our analysis provides critical insights that traditional regression methods cannot offer. We demonstrate that neglecting robust causal inference techniques can lead to inaccurate conclusions about the effect of renewable energy generation on wholesale prices, potentially misguiding analysts and policy makers. While prior studies laid the groundwork for understanding the relationship between renewable energy and electricity prices, our approach offers a new perspective by applying a rigorous analytical and causal lens to uncover the complex, non-linear effects induced by wind and solar energy.

\section*{Relationship between renewable energy penetration and electricity prices}

\begin{figure}[!ht]
    \centering
    \begin{subfigure}[b]{0.48\linewidth}
        \centering
        \includegraphics[width=\linewidth]{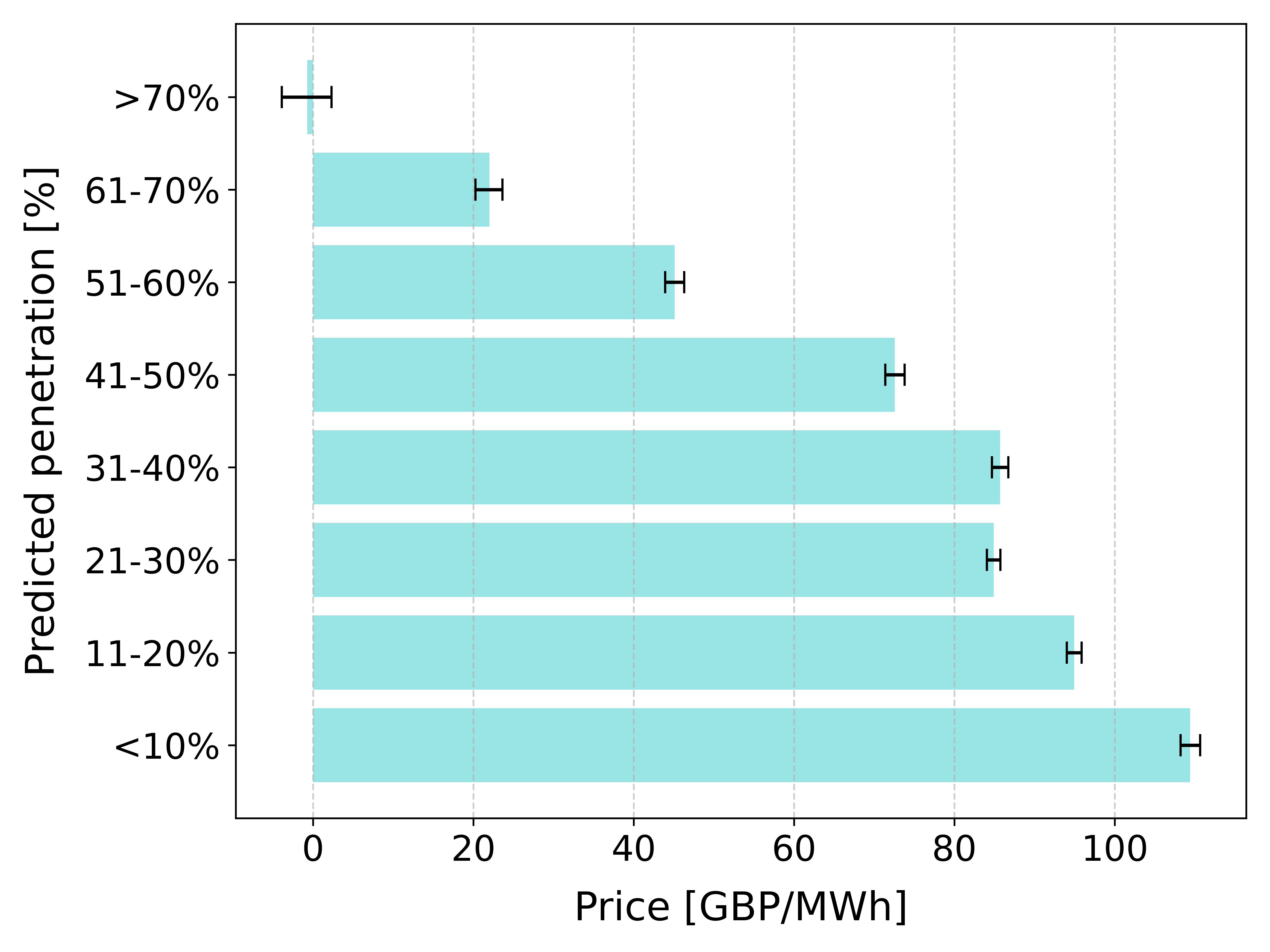}
        \caption{}
        \label{fig:barplot-a}
    \end{subfigure}
    \hspace{.2cm}
    \begin{subfigure}[b]{0.48\linewidth}
        \centering
        \includegraphics[width=\linewidth]{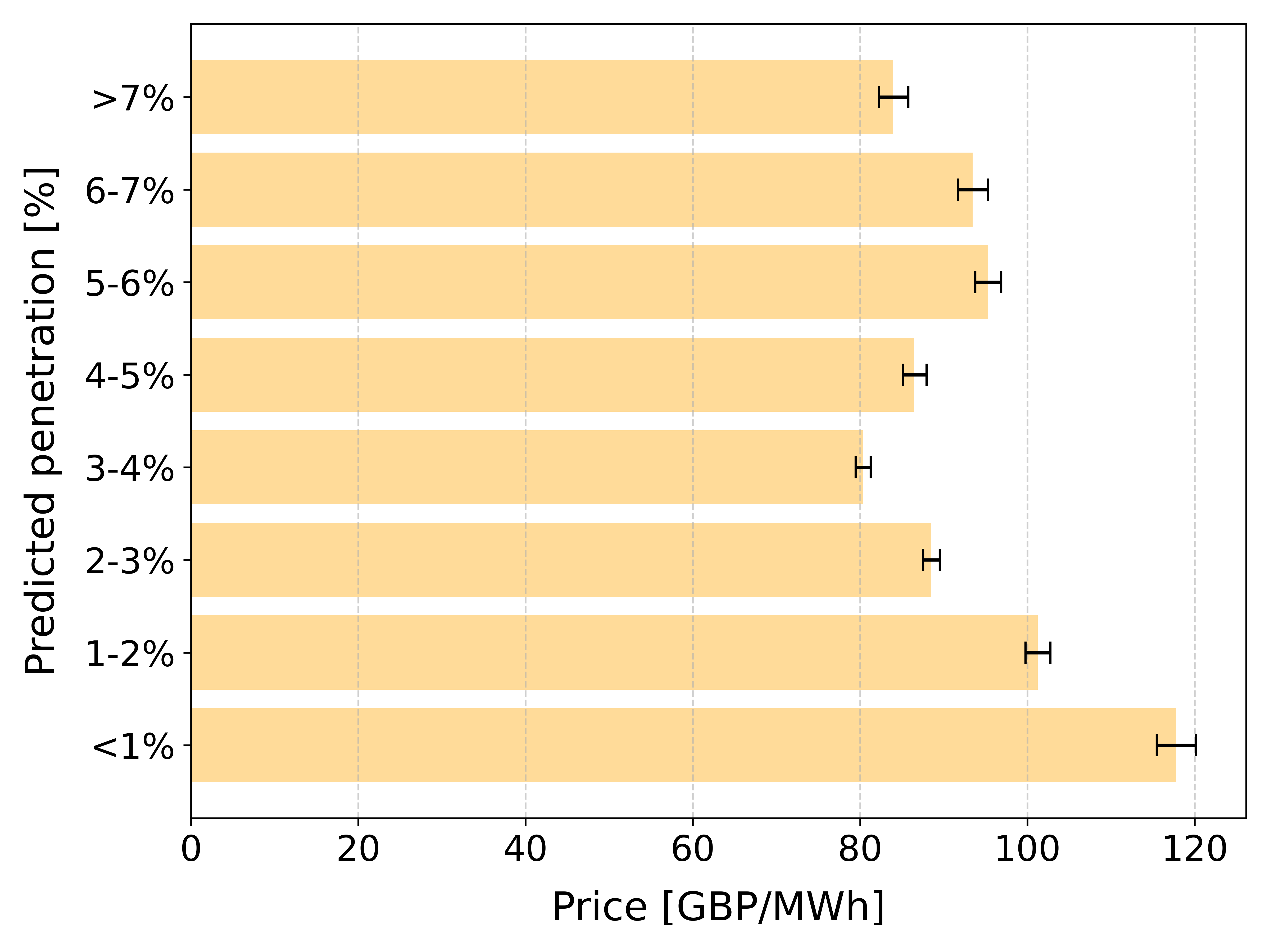}
        \caption{}
        \label{fig:barplot-b}
    \end{subfigure}
    \vspace{1em} 
    \caption{\textbf{Price variation across predicted renewable penetration levels.} Bar plots show the mean APX prices for wind and solar penetration levels, over 2018-2024, with error bars representing associated 95\% confidence intervals. The predicted wind power penetration intervals (a) are defined in 10\% increments, while those for predicted solar power penetration (b) are defined in 1\% increments, reflecting the lower installed solar power capacity in the UK.}
    \label{fig:barplot}
\end{figure}

We first examine the relationship between predicted renewable energy penetration and wholesale electricity prices in the UK over a period from 2018 to 2024. Figure \ref{fig:barplot} presents bar plots of the mean day-ahead APX prices in the UK across varying levels of predicted wind and solar power penetration. For wind power penetration (Figure \ref{fig:barplot-a}), a clear downward trend emerges, where higher penetration levels are associated with lower electricity prices. This aligns with the well-known merit-order effect, as increased predicted renewable energy generation displaces higher-cost conventional generation, reducing market prices. In contrast, for solar power penetration (Figure \ref{fig:barplot-b}), the pattern is more complex. While moderate solar penetration levels are linked to lower prices, an unexpected ``bump'' in the price distribution emerges at around 4–7\% penetration. This counter-intuitive phenomenon may be influenced by external factors such as higher demand during summer months, energy supply disruptions, or other market conditions. Without careful causal analysis, these fluctuations could be mis-attributed solely to solar penetration rather than a broader set of intertwined effects within electricity markets.


\begin{figure}[!ht]
    \centering
    \begin{subfigure}[b]{0.48\linewidth}
        \centering
        \includegraphics[width=\linewidth]{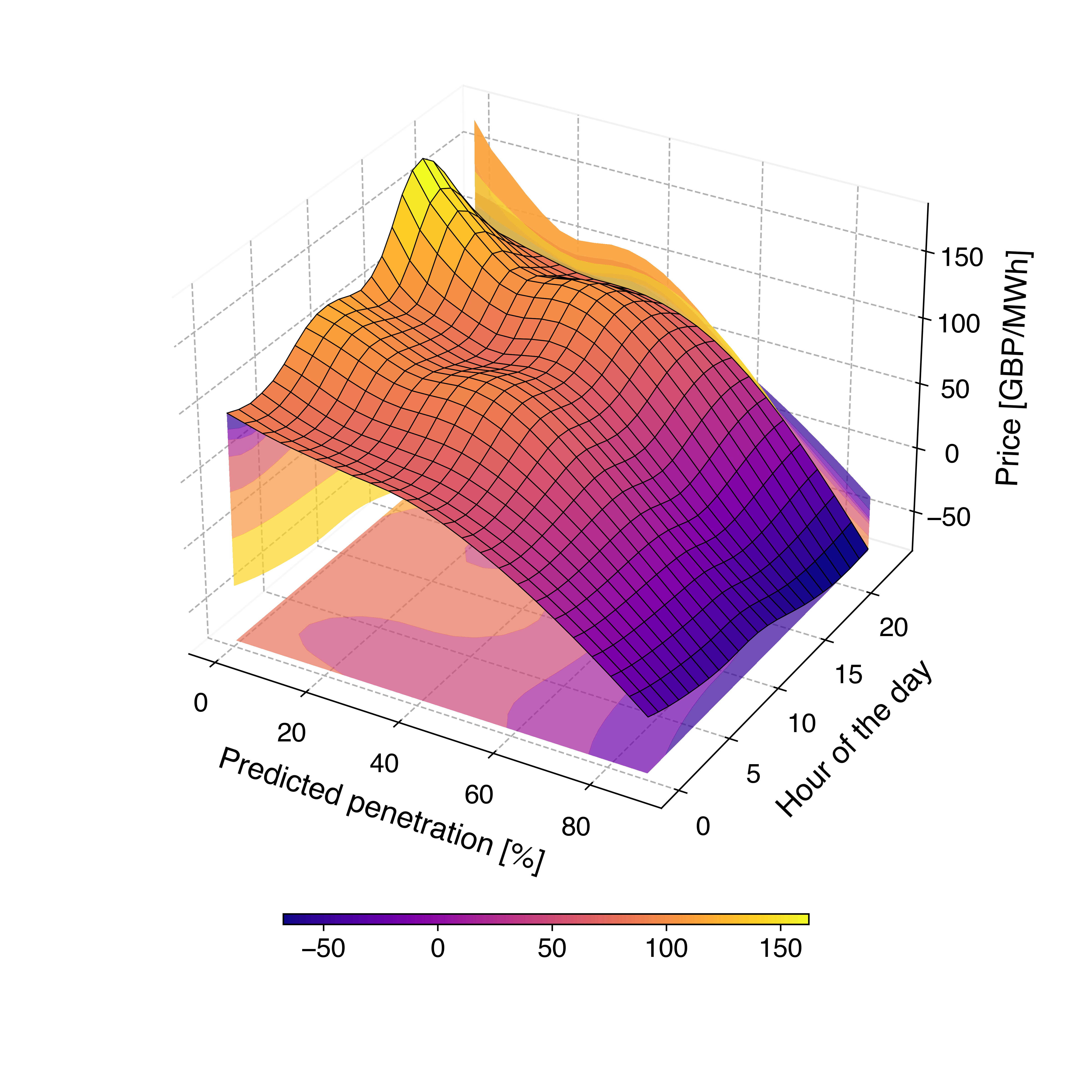}
        \caption{}
        \label{fig:regression-a}
    \end{subfigure}
    \hfill
    \begin{subfigure}[b]{0.48\linewidth}
        \centering
        \includegraphics[width=\linewidth]{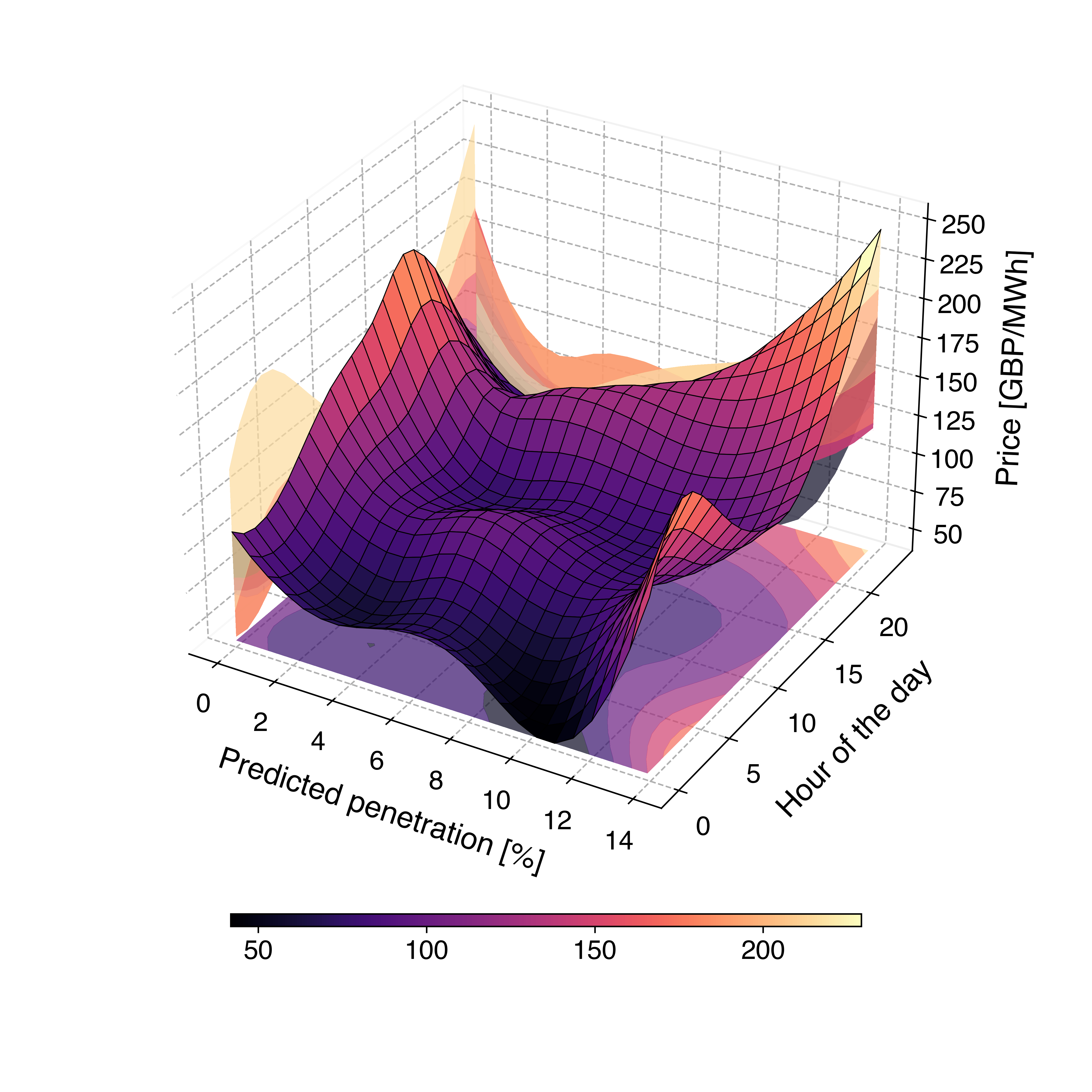}
        \caption{}
        \label{fig:regression-b}
    \end{subfigure}
    \begin{subfigure}[b]{0.48\linewidth}
        \centering
        \includegraphics[width=\linewidth]{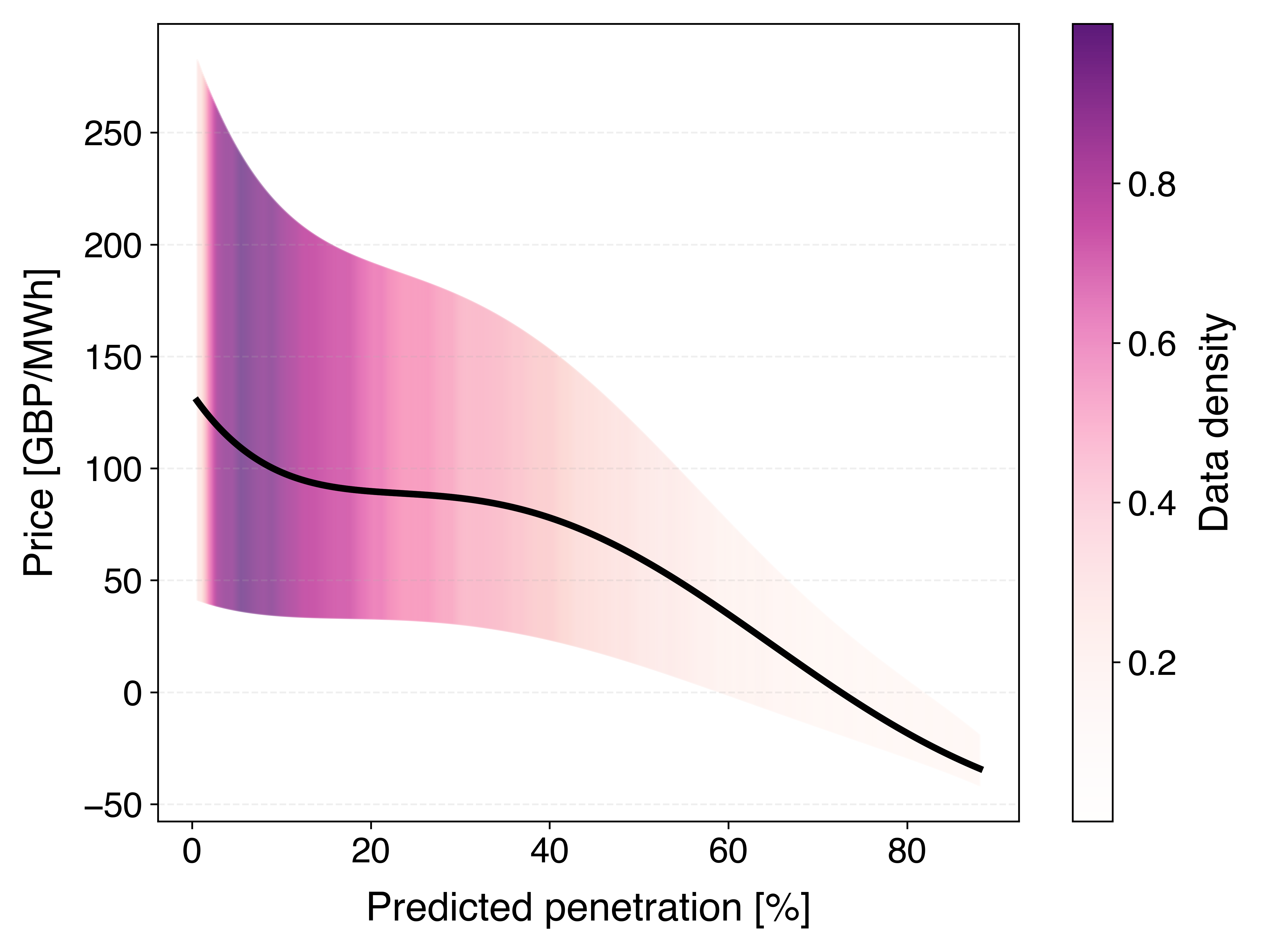}
        \caption{}
        \label{fig:regression-c}
    \end{subfigure}
    \hfill
    \begin{subfigure}[b]{0.48\linewidth}
        \centering
        \includegraphics[width=\linewidth]{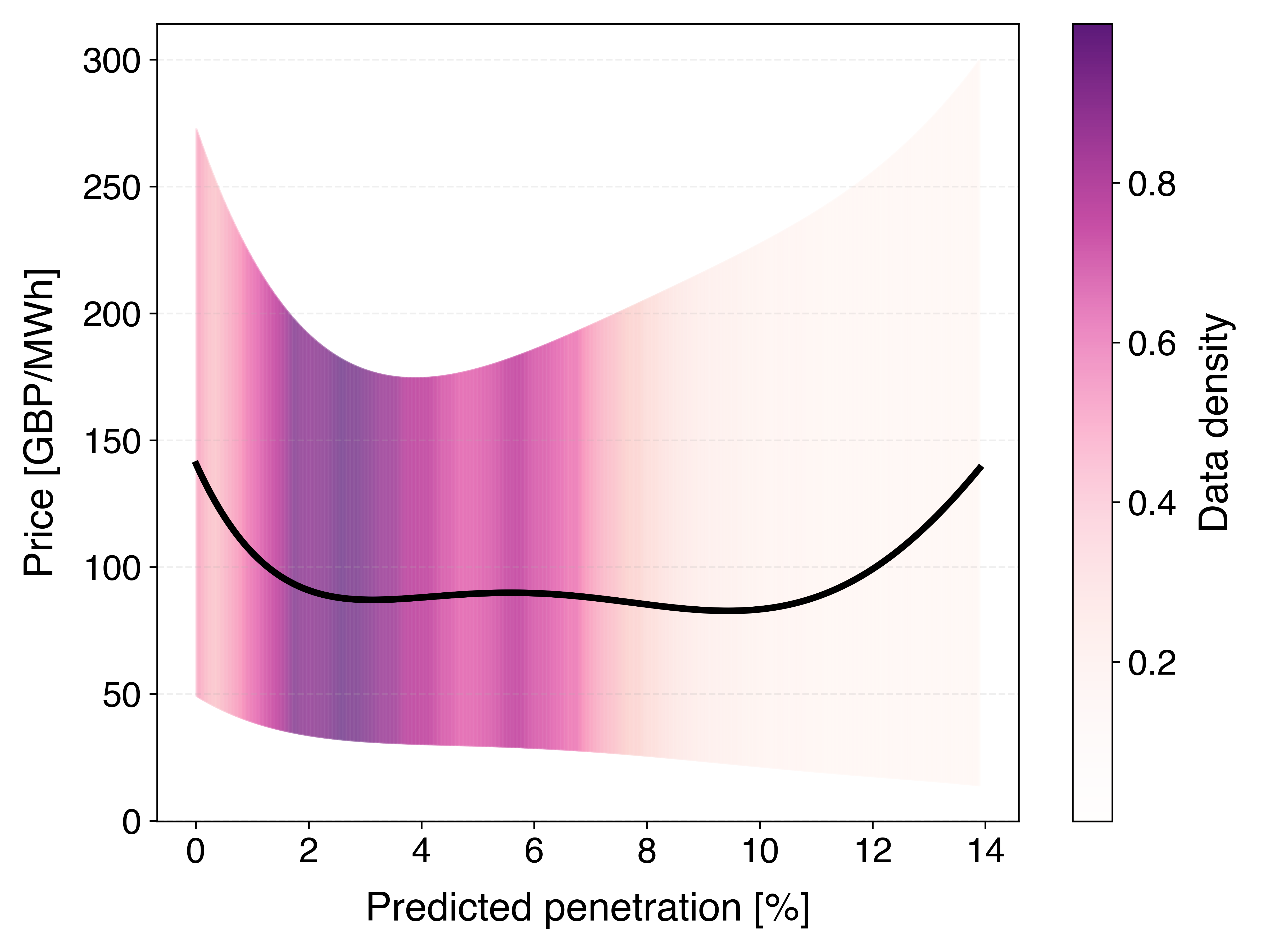}
        \caption{}
        \label{fig:regression-d}
    \end{subfigure}
    \caption{\textbf{Regression analysis of predicted renewable penetration and APX wholesale prices.} The top row shows the mean spot price modelled as a function of the hour of the day, as well as predicted (a) wind power penetration and (b) solar power penetration. The bottom row presents the results of quantile regression models, with the solid line representing the mean spot price, and the shaded region representing central 80\%-coverage intervals, where the colour intensity indicates the data density (in terms of penetration data). These models capture the distributional effects of predicted (c) wind power penetration and (d) solar power penetration, on spot prices.}
    \label{fig:regression}
\end{figure}

To reinforce these initial findings, we performed a regression-based analysis using techniques commonly employed in the literature (using local polynomial (quantile) regression~\cite{jonsson2010market}). Figure \ref{fig:regression} illustrates the relationship between forecast wind and solar power penetration and spot prices, over the same period. For wind power (Figure \ref{fig:regression-a}), the results corroborate earlier insights, showing that higher predicted wind power penetration generally leads to lower wholesale electricity prices. Specifically, as predicted wind power penetration increases, daily price patterns undergo a noticeable transformation: the pronounced peaks during morning and evening hours at low penetration levels are significantly smoothed. This indicates that wind power forecasts not only reduces average prices but also dampen daily price spikes. Results from the quantile regression model (Figure \ref{fig:regression-c}) provide further insights into the distributional effects of wind power generation. With higher levels of predicted wind penetration, the range between the 10\% and 90\% quantiles narrows, suggesting reduced price variability. However, the lighter colour shading at high penetration levels indicates limited data in these regions, cautioning against overgeneralising these findings. 

The analysis of predicted solar power penetration (Figures \ref{fig:regression-b} and \ref{fig:regression-d}) reveals a more complex relationship. As noted earlier (Figure \ref{fig:barplot-b}), higher levels of solar penetration appear to be linked with rising spot prices (which is counter-intuitive). The discrepancy between observed behaviour for wind and solar power also highlights the importance of penetration levels. With wind energy being more pervasive in the UK market, its effect is easier to discern, whereas the impact of solar energy is potentially overshadowed by external factors. These results underscore the limitations of regression-based approaches in disentangling the causal effects of renewable energy penetration. Without accounting for confounding variables, we risk drawing misleading conclusions, such as inferring that higher solar power penetration drives higher wholesale prices. This emphasises the necessity of devising and employing robust causal inference methods to accurately attribute changes in electricity prices to renewable energy penetration.

\section*{Causal impact of renewables on electricity prices}
\begin{figure}[!ht]
    \centering
    \begin{subfigure}[b]{0.48\linewidth}
        \centering
        \includegraphics[width=\linewidth]{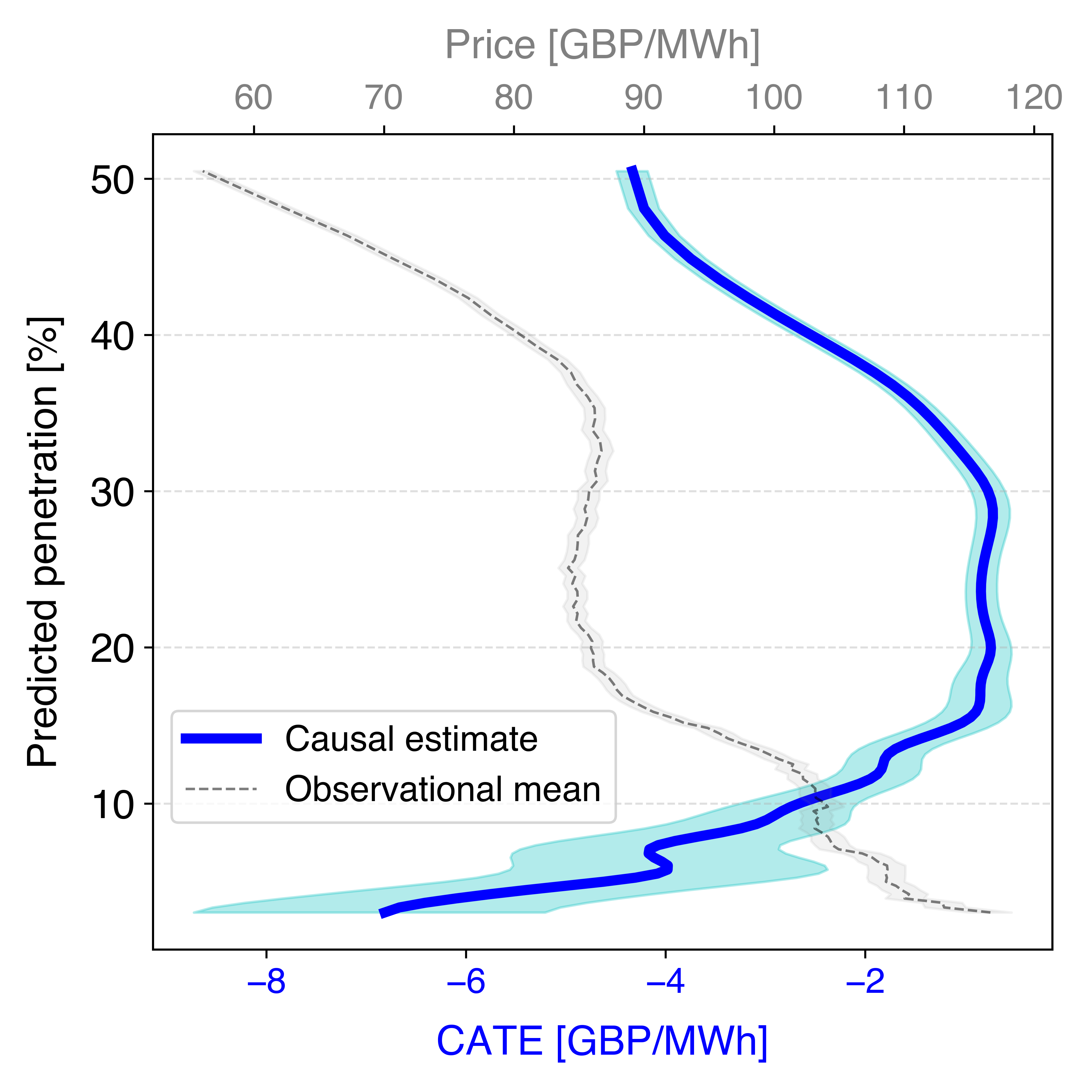}
        \caption{}
        \label{fig:dml-a}
    \end{subfigure}
    \hfill
    \begin{subfigure}[b]{0.48\linewidth}
        \centering
        \includegraphics[width=\linewidth]{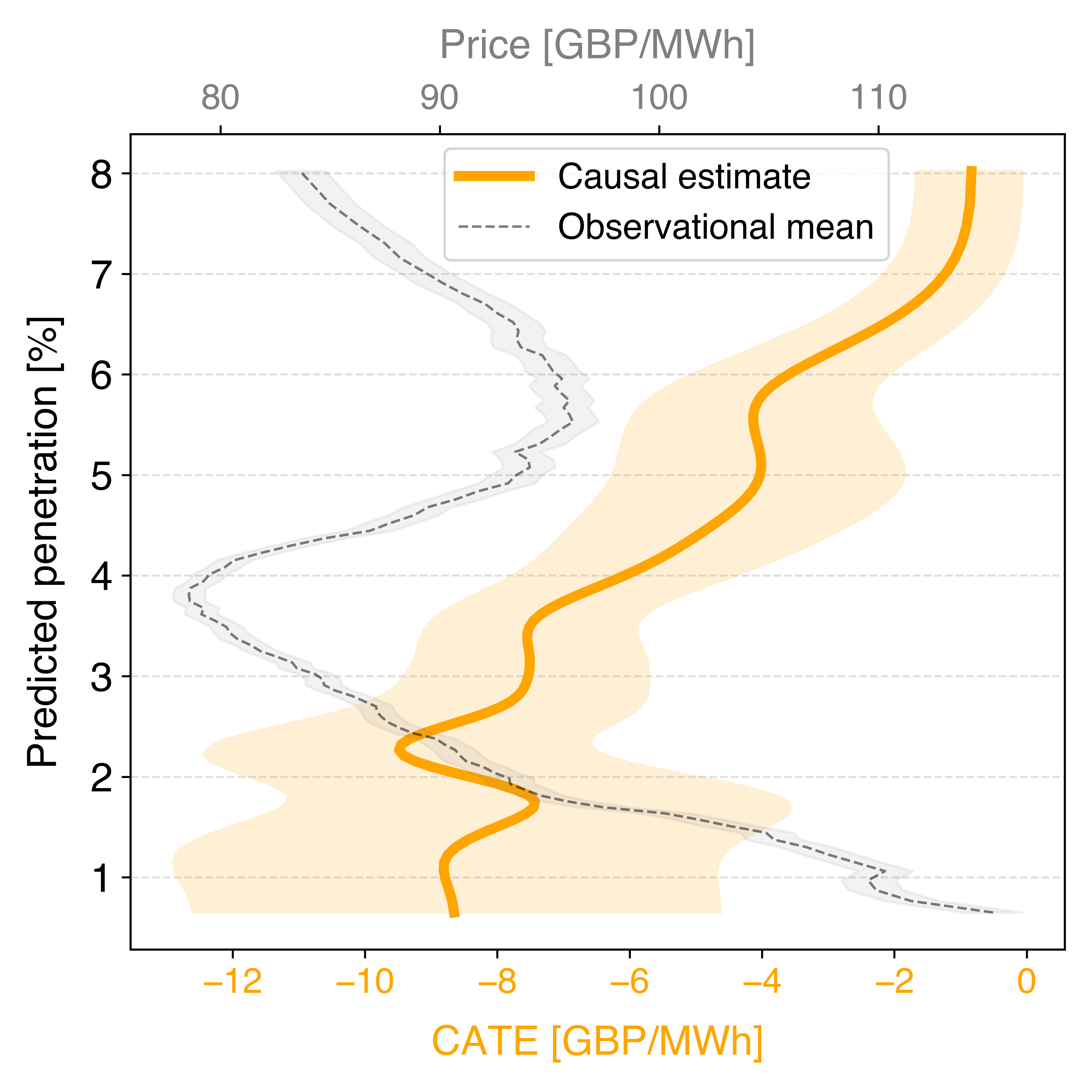}
        \caption{}
        \label{fig:dml-b}
    \end{subfigure}
    \caption{\textbf{Causal impact of predicted renewable power production on electricity prices.} Comparison of observational mean and causal effects of wind (a) and solar (b) power production on wholesale electricity prices. The solid lines represent the non-linear CATE estimates derived using our local partially linear DML framework, capturing the price impact (in GBP/MWh) of a 1 GWh increase in renewable energy generation. The dashed lines show the observational mean trends as a function of renewable penetration levels, illustrating the differences between raw associations and the true causal effects. Shaded areas denote 80\% confidence intervals.}
    \label{fig:dml}
\end{figure}

Understanding the impact of wind and solar energy generation on electricity prices is inherently a causal question: what happens to electricity prices if we predict one more GWh of wind or solar energy? This inquiry goes beyond a simple statistical association; it seeks to uncover the direct, underlying effect of additional renewable energy integration. Historically, experimentation has been central to scientific discovery, providing controlled environments to isolate causes and effects. However, in many real-world environments, such as electricity markets, conducting controlled experiments is neither feasible nor practical. In these cases, deriving causal insights from observational data alone remains one of the most formidable challenges in science and statistics \cite{pearl2009causality, peters2017}. Given the limitations of traditional regression models, we employ an approach based on DML to move beyond simple associations and uncover the true effects of renewable integration. Specifically, our \textit{local partially linear DML} framework enables us to isolate the non-linear impact of wind and solar power production on wholesale electricity prices. This methodology addresses confounding factors that often distort observational analyses, providing a robust foundation for understanding how renewable integration affects market dynamics. Figure \ref{fig:dml} shows the results of our DML framework on the APX prices, for the period 2018-2024 (results for NordPool and intraday prices are included in the supplementary information of the paper, in Figures \ref{fig:dml-nordpool} and \ref{fig:dml-intraday}). The \textit{causal estimate} is the conditional average treatment effect (CATE), which represents the impact of a 1 GWh increase in predicted renewable energy generation on wholesale electricity prices, while controlling for a wide set confounders. Instead, the \textit{observational mean} reflects the raw association between renewable penetration and electricity prices at the same penetration level, without accounting for confounding factors. The full list of potential confounding variables to consider is given in Table \ref{tab:dataset_description} of the supplementary information. It includes variables like varying installed capacity (for wind and solar energy), time of day, time of year, load, gas price, carbon permits, etc. 

The causal analysis reveals distinct effects for wind and solar power. Wind power exhibits a U-shaped causal effect on electricity prices (Figure \ref{fig:dml-a}): the price-reducing effect from wind energy generation is strongest at low penetration levels, diminishes at moderate levels, and increases again at higher penetration levels. This pattern likely reflects the interaction between wind energy generation and the slope of the supply curve in the wholesale electricity markets, at various penetration levels. At low wind power penetration, wind power rapidly displaces the most expensive marginal generators, leading to substantial price reductions. As penetration increases, this effect diminishes temporarily, only to re-emerge as wind displaces more entrenched higher-cost generators. The reduced effect in the moderate range (20-30\%) could arise because the supply curve flattens as gas providers with similar cost structures enter the market, reducing the sensitivity of prices to additional wind power generation. For solar power (Figure \ref{fig:dml-b}), our causal inference approach uncovers a consistent price-reducing effect across all penetration levels. This finding stands in sharp contrast to the observational mean trends, which suggest a ``bump'' in electricity prices at moderate solar penetration (around 5\%). This bump, observed in simpler analyses, likely arises from un-addressed confounding factors such as demand shifts or major events occurring during periods of high solar output. The causal analysis effectively isolates the intrinsic merit-order effect of solar power, demonstrating its role in reducing electricity prices even when confounders might obscure this relationship. Finally, we hypothesise that the effect of solar power on electricity prices might also exhibit a U-shaped pattern, similar to wind, since the effect fundamentally comes from the overall shape of the supply curve in the market. However, this potential trend is not fully observable in our analysis due to the limited attainable penetration of solar power, which reflects the reduced installed solar capacity in the UK. As solar power penetration increases in the future, it is plausible that a more pronounced U-shaped relationship could emerge, driven by dynamics analogous to those seen for wind power. Comparable patterns and causal relationships have been observed while analysing the effects on the NordPool (Figure \ref{fig:dml-nordpool}) and intraday (Figure \ref{fig:dml-intraday}) electricity prices. In the supplementary information of the paper (Figure \ref{fig:dml-v3}), we explored an alternative approach by using predicted renewable penetration not only as a contextual variable but also as the main input to the model. This analysis provides a complementary view by estimating the causal effect of increasing renewable penetration (\%) directly, rather than focusing on energy generation (GWh).

These findings have significant implications for analysts, policy makers and market participants. The U-shaped effect observed in Figure \ref{fig:dml-a} underscores the importance of considering how renewable energy influences market structures across varying levels of market penetration. For solar power, the discrepancy between observational and causal trends underscores the risks of relying on simpler regression methods, which may lead to disproportionately erroneous conclusions about the economic impacts of renewables. By providing an accurate assessment of the true effects of renewable generation, our approach supports better-informed market designs and policy interventions aimed at maximising the benefits from renewable energy integration.

\section*{Evolution of causal effects over time}

\begin{figure}[!ht]
    \centering
    \begin{subfigure}[b]{0.48\linewidth}
        \centering
        \includegraphics[width=\linewidth]{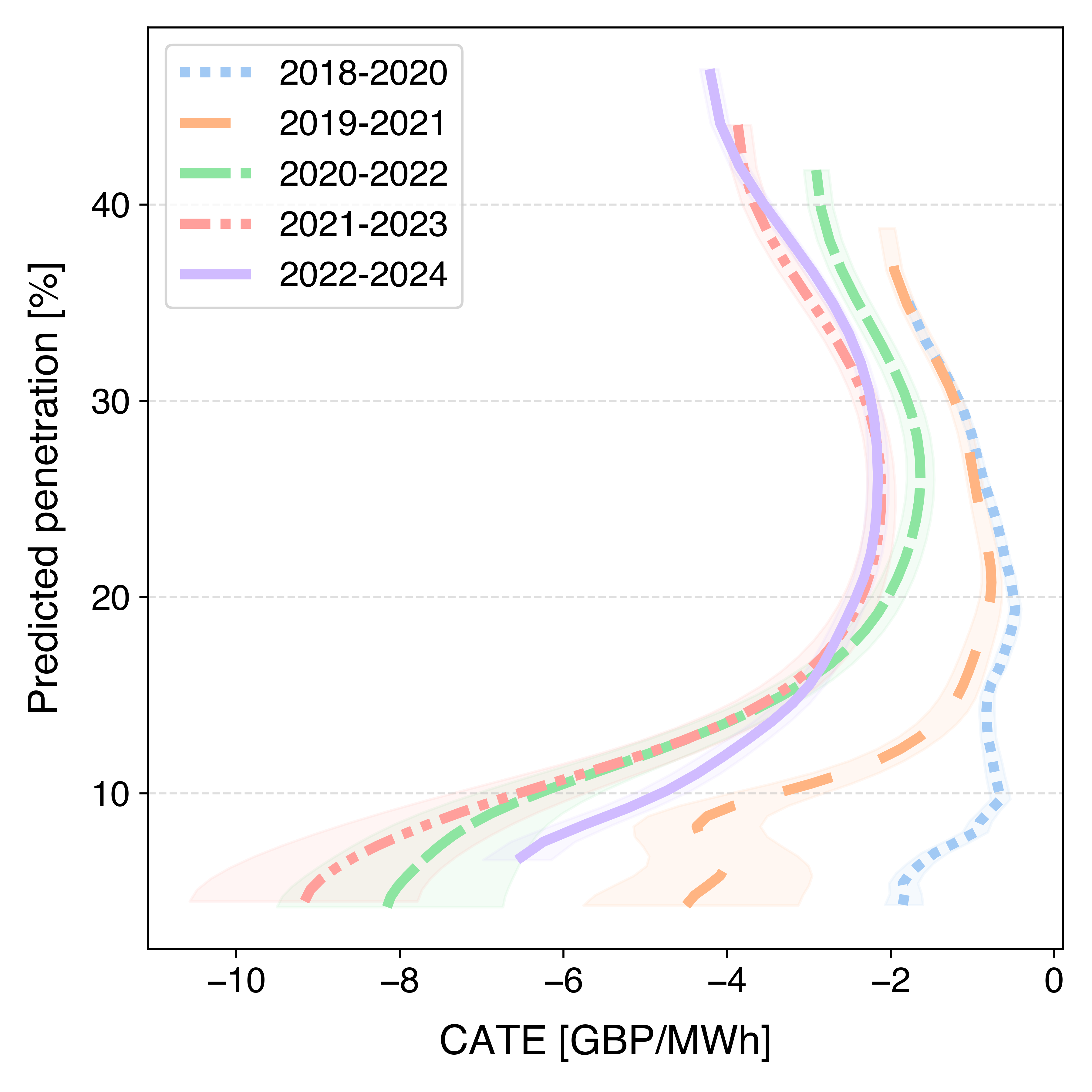}
        \caption{}
        \label{fig:dml-yearly-trend-a}
    \end{subfigure}
    \hfill
    \begin{subfigure}[b]{0.48\linewidth}
        \centering
        \includegraphics[width=\linewidth]{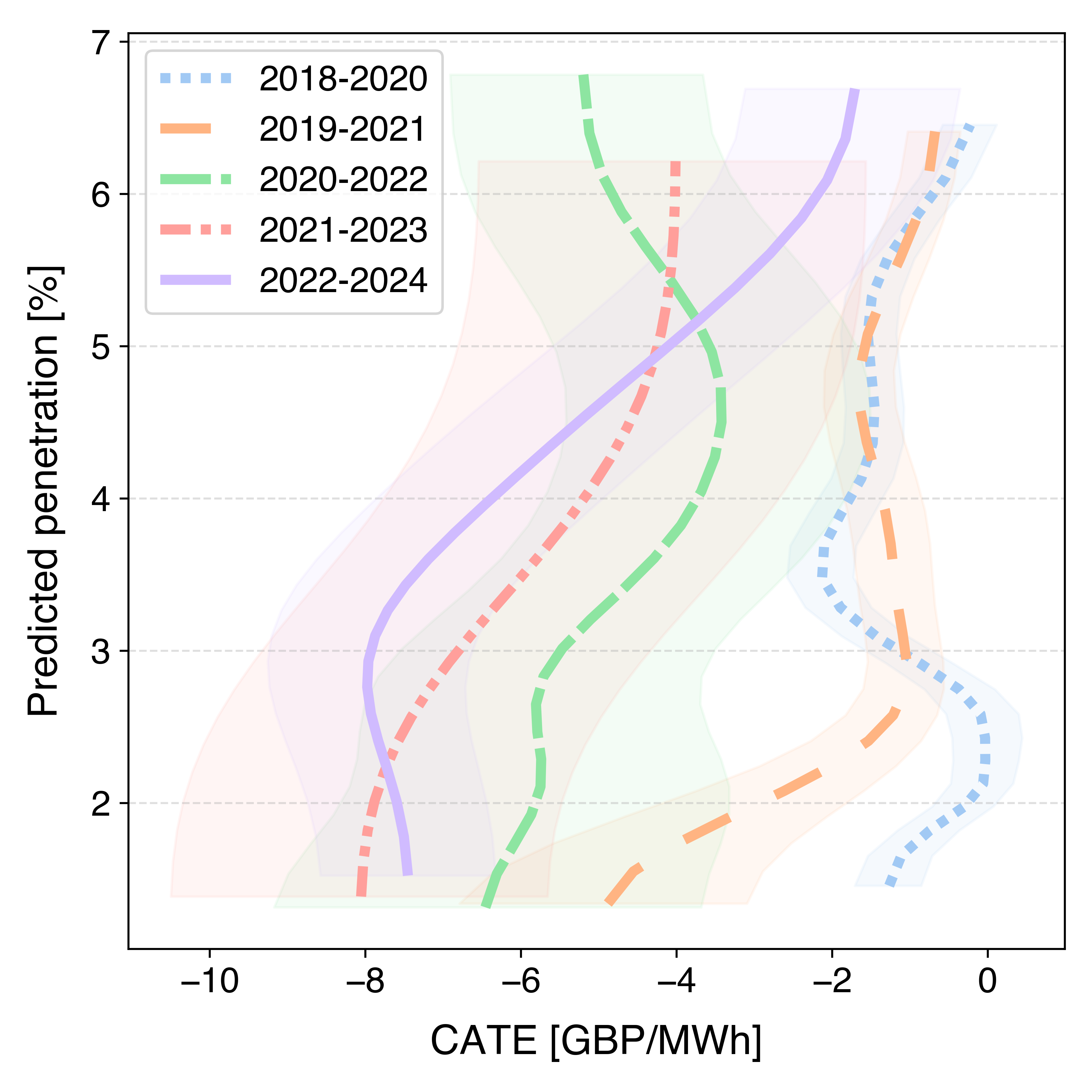}
        \caption{}
        \label{fig:dml-yearly-trend-b}
    \end{subfigure}
    \caption{\textbf{Temporal evolution of the causal impact of renewable power production on electricity prices}. non-linear CATE estimates of (a) wind and (b) solar power production over time, obtained using a sliding window spanning two financial years, with each window containing approximately 35,000 observations. Shaded regions represent 80\% confidence intervals.}
    \label{fig:dml-yearly-trend}
\end{figure}

\noindent
The impact of renewables on electricity prices is not static. It evolves alongside market conditions, technological advancements, and changes in renewable penetration levels driven by additional installed generation capacity. To investigate these dynamics, we analyse the non-linear causal effect over time using a sliding window approach spanning two financial years. This approach provides a temporally resolved view of how the price effects of wind and solar energy have changed over the period 2018-2024, while ensuring a sufficient number of observations to apply our DML framework. The findings in Figure \ref{fig:dml-yearly-trend} reveal a growing impact of renewable energy on electricity prices over the years. In particular, the magnitude of wind’s price-reducing effect (Figure \ref{fig:dml-yearly-trend-a}) has significantly increased in the last four years, reflecting its expanding role in the electricity market and the growing relevance of the merit-order effect. This trend underscores the profound influence of wind power on market outcomes, highlighting its policy implications as governments and regulators aim to integrate higher levels of renewable energy into the grid. The evolution of the effect of solar power on electricity prices (Figure \ref{fig:dml-yearly-trend-b}) also appears to be increasing, indicating that the impact of solar energy may become substantial in a near future with the further deployment of solar power generation capacity in the UK.

\section*{Discussion and conclusions}\label{sec:discussion}
We bridge a critical gap in understanding the causal impact of renewable energy penetration on electricity prices, offering insights that extend beyond regression-based associations only. By applying a robust causal inference framework, we demonstrated how solar power consistently reduces electricity prices, underscoring its fundamental merit-order effect. Similarly, the U-shaped causal effect of wind power highlights the non-linear impact of wind integration, linked to the slope of the overall market supply curve at the equilibrium point, at varying penetration levels. Our analysis further revealed the temporal trends of these causal effects, demonstrating how the influence of renewables on electricity prices has steadily intensified over the period from 2018 to 2024. These findings carry important implications for energy policy makers and market participants. Traditional analyses that rely solely on observational data risk misinterpreting the economic effects of renewables, potentially mis-informing future policy interventions in electricity markets. By contrast, our causal framework provides evidence-based insights that more accurately reflect the true economic benefits of renewables. This evidence underscores the importance of embedding causal methods into the decision-making processes that shape future energy systems.

The implications of our findings extend beyond the UK market only. As many electricity markets transition to higher shares of renewables, the lessons from our analysis become increasingly relevant. The methodology we employ can be adapted to other contexts, providing a versatile tool for understanding how renewable integration impacts market outcomes under diverse conditions. Furthermore, the effects we observe are likely to amplify as renewable energy penetration increases globally, suggesting that the price-reducing effect of renewables will become even more pronounced in markets with higher targets for wind and solar energy. This study also highlights areas for future research. A key question involves similarly assessing the impact of renewables on balancing costs and the growing need for reserves to address variability, uncertainty, and grid constraints. Additionally, exploring interactions between renewable energy penetration and emerging technologies such as long-duration storage and demand-side management, or evaluating the broader social benefits of renewables such as emissions reductions and energy equity through a causal lens would provide a more comprehensive picture of their societal value.

In conclusion, our findings emphasise the importance of grounding energy policy and market interventions in robust causal evidence. As the deployment of wind and solar power generation capacities is poised to strengthen in the future, adopting analysis methodologies able to accommodate the complexities of modern electricity systems will be essential for achieving affordable, sustainable, and reliable power systems. By advancing the causal understanding of renewable energy impact on electricity markets, we contribute to a foundation for smarter policies and a smoother transition to a decarbonised future.

\section*{Methods}\label{sec:methods}

This study employed different layers of analysis to explore the impact of predicted wind and solar power penetration on electricity prices. To provide a baseline using methodologies commonly employed in the literature, we first analysed the relationship between predicted renewable penetration and spot prices using regression analysis. Then, we developed a robust causal inference approach to estimate the evolving and non-linear impact of renewables while controlling for a large set of confounding variables.

\subsection*{Data} \label{subsec:data}
The dataset for this study comprises electricity market data, wind and solar generation forecasts, and various other market-relevant variables for the UK, spanning the period from 2018 to 2024. A comprehensive list of data sources, detailed descriptions of the variables, and the preprocessing steps used to derive additional variables are provided in the supplementary information of the paper (Table \ref{tab:dataset_description}).

\subsection*{Regression-based analysis}
\paragraph{Mean smoothing.} Given the well-established non-linear nature of the relationship between electricity prices, predicted renewable energy penetration, and time of day, we employed locally weighted polynomial regression models \cite{jonsson2010market}. These models assume that although the overall relationship is non-linear, it can be locally approximated by a series of linear models. So, by choosing a set of $m$ fitting points, we can locally estimate $m$ linear models representing wholesale electricity prices as a function of predicted wind or solar penetration and time of the day. For this study, we choose 576 fitting points arranged in a \(24 \times 24\) grid. At each fitting point $\mathbf{x}_u$, given that we have $n$ observations $(\mathbf{z}_i,p_i)$ available, the price $p_i$ is modelled as 
\begin{equation}
    p_i = \boldsymbol{\beta}_u ^\top \mathbf{z}_i + \varepsilon_i, \quad i=1,\hdots,n \, ,
\end{equation} 
where $\boldsymbol{\beta}_u$ represents the vector of regression coefficients tailored specific to the fitting point $\mathbf{x}_u$, $\mathbf{z}_i$ comprises a vector of input features, and $\epsilon_i$ denotes the error term (centred, and with finite variance). While with a higher number of fitting points we could adequately approximate non-linear functions with simple first-order models, using polynomial features enables the modelling of more complex relationships. If we let $r_{i}$ and $h_{i}$ represent the normalised renewable power level (wind or solar power) and the hour of the day, we can obtain a second-order model by including quadratic and interaction terms, after expanding each data point $\mathbf{x}_i = [r_i \,\, h_i]^\top$ to $\mathbf{z}_i = [1 \,\, r_{i} \,\,  h_{i} \,\,  r_{i}^2 \,\,  r_{i}h_{i}\,\,  h_{i}^2]^\top$. Locally weighted polynomial regression then estimates the model coefficients $\boldsymbol{\beta}_u$ by weighting observations based on their proximity to the fitting point (using a distance $d(\mathbf{x}_u,\mathbf{x}_i)$). This is achieved through a kernel function that assigns higher weights to nearby observations, reducing influence as the distance increases. Here, we used the tri-cube weight function given by
\begin{equation} \label{eq:weight}
    w_{u,i} = 
    \begin{cases}
        (1 - d(\mathbf{x}_u,\mathbf{x}_i)^3)^3 \, , & \text{if } d(\mathbf{x}_u,\mathbf{x}_i) < 1 \\
        0  \, , & \text{otherwise}
    \end{cases} \, ,
\end{equation}
\noindent where $d$ is chosen as the normalised Euclidian distance between an observation $\mathbf{x}_i$ and the fitting point at hand $\mathbf{x}_u$, i.e., $d(\mathbf{x}_u,\mathbf{x}_i)=||\mathbf{x}_i-\mathbf{x}_u||_2/h_u$. An adaptive bandwidth $h_u$, for each fitting point $\mathbf{x}_u$, can be used to ensure that observations beyond a certain distance from the current point exert no influence on parameter estimation. We calculated $h_u$ using the 30th percentile of distances from all data points to a given fitting point, accommodating to local data density and variability. At each fitting point, the estimated parameter vector $\hat{\boldsymbol{\beta}}_u$ is obtained by minimising the weighted sum of squared residuals, as in 
\begin{equation}
    \hat{\boldsymbol{\beta}}_u = (\mathbf{Z}^{\top}\mathbf{W}_u \, \mathbf{Z})^{-1}\mathbf{Z}^{\top} \mathbf{W}_u \, \mathbf{p} \, ,
\end{equation} 
where $\mathbf{W}_u$ is a diagonal matrix containing the weights $w_{u,i}$, $\mathbf{Z}$ is the design matrix expanded to the quadratic model form (i.e., whose rows are all input feature observations $\mathbf{z}_i^\top$), and $\mathbf{p}$ is the vector of observed response values $p_i$. This methodology effectively captures local variations by emphasising the influence of nearby observations, thus providing a detailed and smooth representation of the relationship between wind power production, time of day, and wholesale prices.

\paragraph{Quantile modelling.} Understanding the relationship between predicted renewable penetration, time of the day, and mean wholesale prices provides interesting insights on the behaviour of the market. However, to cope with the inherent uncertainty of renewable generation, it is equally important to describe the distribution of data around the mean trend. Using quantile regression, we can get insights into the conditional distribution of the price, indicating the value below which a certain proportion of observations fall for a given quantile with nominal level $q$. For example, if $q=0.9$, the quantile regression line will show the value below which 90\% of the observed data points lie. Locally weighted polynomial regression can be extended to model quantiles by minimising a cost function that differently weights prediction errors depending on whether they fall below or above the specified quantile. This approach captures the asymmetric risk associated with underestimations or overestimations. The model parameters for a specific quantile $q$ are determined by
\begin{equation} \label{eq:quantile}
    \hat{\boldsymbol{\beta}}_{u,q} = \argmin_{\boldsymbol{\beta}} \sum_{i=1}^n w_{u,i} 
    \, \rho_q (p_i - \mathbf{z}_i^\top \boldsymbol{\beta}) \, ,
\end{equation}
where $p_i$ represents the observed price, $\mathbf{z}_i$ is the $i$-th observation expanded to the model form and $w_i$ is its weight (computed using \eqref{eq:weight}), $\boldsymbol{\beta}$ is the vector of coefficients to be estimated, and $\rho_q$ is the quantile loss function defined as $\rho_q(\varepsilon) = \varepsilon \, (q - \mathbf{1}\{\varepsilon < 0\})$, with $\mathbf{1}\{ \varepsilon < 0\}$ being an indicator function that takes the value 1 if the residual $e$ is negative and 0 otherwise, and $q$ represents the nominal level of the quantile of interest (e.g., 0.1 for the 10\% quantile and 0.9 for the 90\% quantile). Unlike the case for mean modelling, there is no closed-form solution for the parameter vector $\hat{\boldsymbol{\beta}}_{u,q}$, which is estimated at each fitting point using an iterative gradient-based optimisation method to minimise the quantile loss function in \eqref{eq:quantile}. This methodology provides a detailed representation of the behaviour of prices, beyond average effects, for varying levels of predicted wind power production or penetration.

\subsection*{Causal inference using double machine learning (DML)} \label{subsec:DML}
Traditional regression models may fall short in disentangling causation from confounding influences, especially when the data involve complex, non-linear interactions or high-dimensional confounders like fuel prices, demand fluctuations, and seasonal patterns. The DML framework \cite{chernozhukov2018double} leverages machine learning algorithms to flexibly model the relationships between confounders, treatments, and outcomes. This approach enables the estimation of causal effects in the presence of complex, high-dimensional data while mitigating the spurious effects introduced by confounders.  Within the partially linear DML framework \cite{fuhr2024estimating}, we assume that the response $p$ (here, the electricity price) is a function of the treatment $t$ (e.g., predicted renewable energy penetration) and other confounding variables $\mathbf{a}$, as in
\begin{equation}
    p_i = \beta t_i + f(\mathbf{a}_i) + \varepsilon_i, \quad i=1,\hdots,n \, ,
\end{equation}
where $\beta$ represents the (linear) effect of $t$ on the outcome $p$, $f$ is a potentially non-linear function of the confounders $\mathbf{a}$, and $\varepsilon$ is an error term (centred, and with finite variance). Similarly, we assume that the treatment variable itself can be modelled as a function of the confounders, or a subset of them, i.e.,
\begin{equation}
    t_i = g(\mathbf{a}_i) + \eta_i, \quad i=1,\hdots,n \, ,
\end{equation}
where $g$ represents the modelled relationship between $t$ and $\mathbf{a}$, and $\eta$ is an error term (centred, and with finite variance). The DML framework is designed to isolate the causal effect of the treatment variable $\beta$ by adjusting for confounding influences in two stages. First, we estimate the nuisance parameters using two machine learning models to approximate the functions $f$ and $g$. Then, we use the trained models to ``remove'' the influence of confounders from both the treatment and the response variables. Then, it regresses the residualised response $\tilde{p}_i = p_i - f(\mathbf{a}_i)$, $i=1,\hdots,n$, on the residualised treatment $\tilde{t}_i = t_i - g(\mathbf{a}_i)$, $i=1,\hdots,n$, to estimate $\beta$. Here, $\beta$ coincides with the \emph{average treatment effect} (ATE), or average causal effect, which measures the overall impact of a treatment on the outcome across the entire population. Formally, for a continuous treatment variable $t$, the ATE is defined as
\begin{equation}
    \text{ATE} = \mathbb{E}[p \mid do(t+1)] - \mathbb{E}[p \mid do(t)] \, ,
\end{equation}
where $p | do(t)$ represent the potential outcome with treatment $t$ and $p | do(t+1)$ the potential outcome after a unit increase in $t$. ATE can be interpreted as the expected increase in the response $p$, resulting from a unit increase in the treatment $t$. The relationship is hence linear and not conditional on any contextual variable.

\subsection*{Local partially linear DML}
While the partially linear DML framework provides reliable causal estimates under the assumption of linear treatment effects, it may not fully capture the complexities of the electricity market, where the effect of renewables may vary based on the penetration level. To address this issue, we extend the standard DML approach by introducing a local partially linear DML framework based on a boxcar kernel. Rather than estimating a global ATE across the entire dataset, we assume that the relationship between treatment (renewable energy generation) and outcome (electricity prices) can be locally approximated using a partially linear model within small, predefined subsets of the data. This enables us to estimate the so-called \emph{conditional average treatment effect} (CATE), which can vary depending on the segment of the data under consideration. Within this context, the boxcar kernel acts as a window function to isolate a subset of the data around each point of interest (e.g., for a given level of penetration, time). The idea is similar in essence to the locally weighted polynomial regression models used previously. For a conditioning (and possibly multidimensional) variable $\mathbf{x}$ and a fitting point $\mathbf{x}_u$, the boxcar kernel is defined as
\begin{equation}
    K_u(\mathbf{x}) =
    \begin{cases}
        1, & \text{if } ||\mathbf{x} - \mathbf{x}_u||_2 \leq \dfrac{h}{2} \\
        0, & \text{otherwise}
    \end{cases} \, ,
\end{equation}
where $h$ is the width of the window. The kernel takes the value 1 within the bandwidth parameter $h$ around $\mathbf{x}_u$ and 0 otherwise. This kernel is applied to partition the data into smaller, rectangular regions where the partially linear model is assumed to hold, yielding a subset $\mathcal{W}(\mathbf{x}_u) = \{(\mathbf{x}_i,\mathbf{a}_i,t_i,p_i), \, K_u(\mathbf{x}_i)=1 \}$. For each segment of the data within the window, we estimate the CATE by applying the partially linear DML method. Formally, at each fitting point $\mathbf{x}_u$ (i.e., a given predicted wind penetration level), we define the local CATE as
\begin{equation}
    \text{CATE}(\mathbf{x}_u)  = \mathbb{E}[p \mid do(t+1), \, \mathbf{x} \in \mathcal{W}(\mathbf{x}_u) ] - \mathbb{E}[p \mid do(t), \, \mathbf{x} \in \mathcal{W}(\mathbf{x}_u) ] \, ,
\end{equation}
where $\mathcal{W}(\mathbf{x}_u)$ is the subset of data within the window centred around the fitting point $\mathbf{x}_u$, and the expectation is taken within this window. We refer to $\mathcal{W}(\mathbf{x}_u)$ as the contextual subset of data. The procedure follows two main steps, similar to the standard DML, but with the additional local partitioning. For each contextual subset $\mathcal{W}(\mathbf{x}_u)$, we use machine learning models to estimate the nuisance models $f$ and $g$, and residualise both the treatment variable $t$ and the outcome $p$. This removes the confounding effects, leaving us with the contextual residualised variables $\tilde{t}_u$ and $\tilde{p}_u$, with vectors of values $\tilde{\mathbf{t}}_u$ and $\tilde{\mathbf{p}}_u$ that corresponds to the contextual subset $\mathcal{W}(\mathbf{x}_u)$ . The variables used for residualising wind power production, solar power production, and spot prices are listed in Table \ref{tab:dataset_description} of the supplementary information. Then, at each window we estimate the CATE with
\begin{equation}
   \hat{\beta}_u = \text{OLS}(\tilde{\mathbf{p}}_u,\tilde{\mathbf{t}}_u \mid \mathbf{x} \in \mathcal{W}(\mathbf{x}_u)) = (\tilde{\mathbf{t}}_u^\top \tilde{\mathbf{t}}_u)^{-1} \, \tilde{\mathbf{t}}_u^\top \tilde{\mathbf{p}}_u \, .
\end{equation}

We use a bootstrap procedure for each contextual subset $\mathcal{W}(\mathbf{x}_u)$ to ensure robust results and to capture the distributional uncertainty of our estimates. Details of the estimation procedure are provided in Algorithms \textcolor{blue}{1} and \textcolor{blue}{2} in the supplementary information for this paper. Additionally, a Gaussian filter was applied to reduce the noise in the CATE estimates and to provide a smoother representation of the non-linear effects. The individual CATE estimates prior to smoothing are presented in the supplementary information (Figures \ref{fig:dml-scatter} and \ref{fig:dml-scatter-yearly}).

\section*{Data availability}
The data utilised in this study is available at \href{https://github.com/dcacciarelli/market-impact-renewables/tree/main/data}{https://github.com/dcacciarelli/market-impact-renewables/tree/main/data}.

\section*{Code availability}
The Python scripts to replicate all the results presented in the paper are available at \href{https://github.com/dcacciarelli/market-impact-renewables}{https://github.com/dcacciarelli/market-impact-renewables}.



\section*{Author contributions statement}
D.C. developed the methodology, conducted the analysis, and drafted the manuscript. P.P. provided continuous supervision in methodology development, experimental design, result analysis, while also contributing to the writing of the manuscript. F.P. and D.D. supported data sourcing and preprocessing and contributed to the result analysis. L.B. conceptualised the initial idea and provided feedback throughout the project. All authors reviewed the manuscript and accepted its submission for potential publication.

\section*{Competing interests}
The authors declare no competing interests.

\clearpage
\section*{Supplementary information}

\vspace{1cm}



\renewcommand{\thefigure}{S\arabic{figure}}
\renewcommand{\thetable}{S\arabic{table}}
\renewcommand{\tablename}{Supplementary Table}

\setcounter{figure}{0}
\setcounter{table}{0}




\subsection*{Supplementary notes}

\vspace{.5cm}

\subsubsection*{Partially linear DML}
The DML framework \cite{chernozhukov2018double} is a powerful approach that integrates machine learning with econometric techniques to provide unbiased causal estimates even when the data is characterised by nonlinear relationships among numerous confounders. In the context of DML, one of the most commonly encountered setup is the partially linear one \cite{fuhr2024estimating}.

\begin{algorithm}
\caption{Partially linear DML}
\label{alg:dml-partially-linear}
\begin{algorithmic}[1]
\State \textbf{Input:} dataset \( D = \{(\mathbf{a}_i, t_i, p_i)\}_{i=1}^n \), number of folds \( K \)
\State \textbf{Output:} linear ATE estimate \( \hat{\beta} \)

\State Randomly partition the dataset into \( K \) folds. \Comment{Split the data}
\For{each fold \( k = 1, \ldots, K \)} \Comment{Train predictive models}
    \State Train a model \( \hat{f}_{-k} \) using \( K-1 \) folds to predict the response \( p \) from the confounders \( \mathbf{a} \).
    \State Train a model \( \hat{g}_{-k} \) using \( K-1 \) folds to predict the treatment \( t \) from the confounders \( \mathbf{a} \).
    \State Use \( \hat{f}_{-k} \) and \( \hat{g}_{-k} \) to predict \( \mathbf{p} \) and \( \mathbf{t} \) in the held-out fold \( k \).
    \State Compute residuals \Comment{Orthogonalisation}
    \begin{align*}
    \tilde{\mathbf{p}}_k &= \mathbf{p}_k - \hat{f}_{-k}(\mathbf{A}_k) \\
    \tilde{\mathbf{t}}_k &= \mathbf{t}_k - \hat{g}_{-k}(\mathbf{A}_k)
    \end{align*}
    \State Regress \( \tilde{\mathbf{p}}_k \) on \( \tilde{\mathbf{t}}_k \) using OLS \Comment{Effect estimation}
    \begin{equation*}
    \hat{\beta}_k = \text{OLS}(\tilde{\mathbf{p}}_k, \tilde{\mathbf{t}}_k) = (\tilde{\mathbf{t}}_k^\top \tilde{\mathbf{t}}_k)^{-1} \, \tilde{\mathbf{t}}_k^\top \tilde{\mathbf{p}}_k
    \end{equation*}
\EndFor
\State Compute the final estimate as the mean:
\begin{equation*}
\hat{\beta} = \frac{1}{K} \sum_{k=1}^K \hat{\beta}_k
\end{equation*}

\end{algorithmic}
\end{algorithm}

The main steps of DML-based effect estimation for the partially linear case are reported in Algorithm \ref{alg:dml-partially-linear}, where $\mathbf{a}_i$ represents the $i$th vector of confounding variables, $t_i$ is the value of the treatment for the $i$th observation, and $p_i$ is the corresponding price value. The function $f$ represents the modelled relationship between $p$ and $\mathbf{a}$, while $g$ represents the modelled relationship between $t$ and $\mathbf{a}$. It should be noted that the models trained in steps 5 and 6 of Algorithm \ref{alg:dml-partially-linear} can be any machine learning model, allowing for flexibility in capturing complex relationships between variables. In our implementation, we employed LightGBM regression models \cite{ke2017lightgbm}.

\clearpage
\subsubsection*{Locally partially linear DML}
Our \textit{locally partially linear DML} approach makes use of a boxcar kernel to estimate local effects (CATEs), instead of a global ATE. The key steps of the estimation procedure are described in Algorithm \ref{alg:dml-locally-linear}, where $\mathbf{x}$ is the conditioning variable used to contextualise the estimation of the causal effect. In our implementation, we used kernels with a size $h$ of 10,000 observations, and a step size $s$ of 1,000 observations. The smoothing parameter of the Gaussian filter was set to 1.5 times the standard deviation of the mean effect estimates. The number of bootstrap replications within each boxcar kernel was set to 100.

\begin{algorithm}
\caption{Locally partially linear DML}
\label{alg:dml-locally-linear}
\begin{algorithmic}[1]
\State \textbf{Input:} dataset \( D = \{(\mathbf{x}_i, \mathbf{a}_i, t_i, y_i)\}_{i=1}^n \), kernel size \( h \), step size \( s \), number of bootstrap replications \( B \), number of folds \( K \)
\State \textbf{Output:} nonlinear CATE estimate

\For{each fitting point \( \mathbf{x}_u \)}
    \State Select subset \( \mathcal{W}(\mathbf{x}_u) = \{(\mathbf{x}_i,\mathbf{a}_i,t_i,p_i), \, ||\mathbf{x} - \mathbf{x}_u||_2 \leq \dfrac{h}{2}\} \) \Comment{Contextual subset of the data} 
    \For{each bootstrap iteration \( b = 1, \ldots, B \)}
        \State Draw a resampled dataset \( \mathcal{W}^b(\mathbf{x}_u) \) from \( \mathcal{W}(\mathbf{x}_u) \) with replacement \Comment{Bootstrap resampling}
        \State Randomly partition the dataset into \( K \) folds. \Comment{Split the data}
        
        \For{each fold \( k = 1, \ldots, K \)} \Comment{Train predictive models}
            \State Train a model \( \hat{f}_{-k}^b \) using \( K-1 \) folds to predict the response \( p \) from the confounders \( \mathbf{a} \).
            \State Train a model \( \hat{g}_{-k}^b \) using \( K-1 \) folds to predict the treatment \( t \) from the confounders \( \mathbf{a} \).
            \State Use \( \hat{f}_{-k}^b \) and \( \hat{g}_{-k}^b \) to predict \( \mathbf{p} \) and \( \mathbf{t} \) in the held-out fold \( k \).
            \State Compute residuals \Comment{Orthogonalisation}
            \begin{align*}
            \tilde{\mathbf{p}}_k^b  &= \mathbf{p}_k^b  - \hat{f}_{-k}^b (\mathbf{A}_k^b ) \\
            \tilde{\mathbf{t}}_k^b  &= \mathbf{t}_k^b  - \hat{g}_{-k}^b (\mathbf{A}_k^b )
            \end{align*}
            \State Regress \( \tilde{\mathbf{p}}_k^b  \) on \( \tilde{\mathbf{t}}_k^b  \) using OLS \Comment{Effect estimation}
            \begin{equation*}
            \hat{\beta}_k^b(\mathbf{x}_u) = \text{OLS}(\tilde{\mathbf{p}}_k^b , \tilde{\mathbf{t}}_k^b ) = (\tilde{\mathbf{t}}_k^{b\top} \tilde{\mathbf{t}}_k^b )^{-1} \, \tilde{\mathbf{t}}_k^{b \top} \tilde{\mathbf{p}}_k^b
            \end{equation*}
        \EndFor
        \State Compute the mean estimate for the current bootstrap sample \( \mathcal{W}^b(\mathbf{x}_u) \) as:
        \begin{equation*}
        \hat{\beta}^b(\mathbf{x}_u)  = \frac{1}{K} \sum_{k=1}^K \hat{\beta}_k^b(\mathbf{x}_u)
        \end{equation*}
    \EndFor
    \State Compute the mean estimate for the current kernel \( \mathcal{W}(\mathbf{x}_u) \) as:
        \begin{equation*}
        \hat{\beta}(\mathbf{x}_u)  = \frac{1}{B} \sum_{b=1}^B \hat{\beta}^b(\mathbf{x}_u)
        \end{equation*}
\EndFor
\State Aggregate \( \hat{\beta}(\mathbf{x}_u) \) across all kernel centres \( \mathbf{x}_u) \) to construct the nonlinear CATE estimate
\State Smooth the aggregated CATE estimates to reduce noise using a Gaussian filter \Comment{Optional}
\end{algorithmic}
\end{algorithm}

\clearpage
\subsection*{Supplementary figures}

\vspace{.5cm}

\subsubsection*{Results from NordPool and intraday markets}

\vspace{.3cm}

\begin{figure}[!ht]
    \centering
    \begin{subfigure}[b]{0.48\linewidth}
        \centering
        \includegraphics[width=\linewidth]{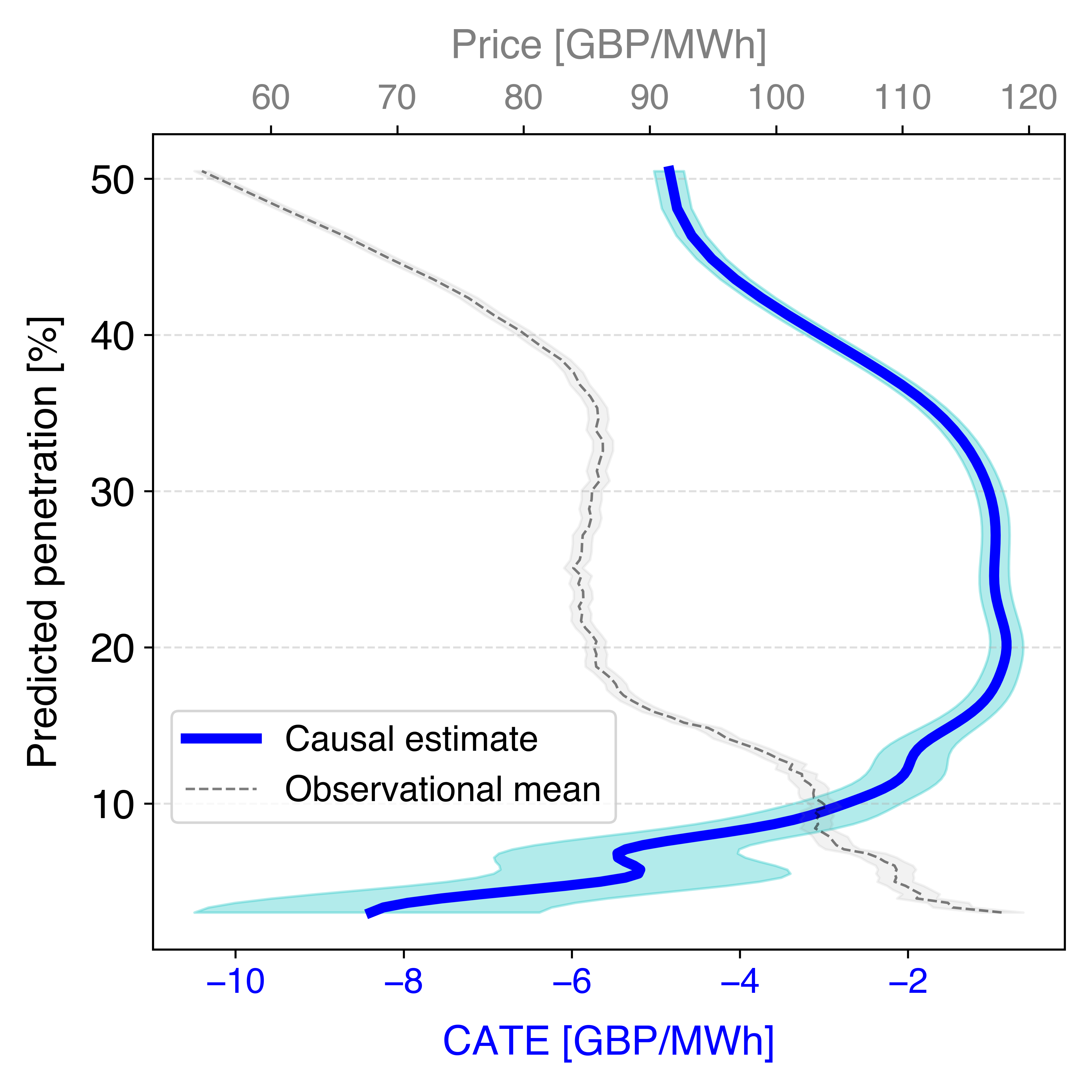}
        \caption{}
        \label{fig:dml-nordpool-a}
    \end{subfigure}
    \hfill
    \begin{subfigure}[b]{0.48\linewidth}
        \centering
        \includegraphics[width=\linewidth]{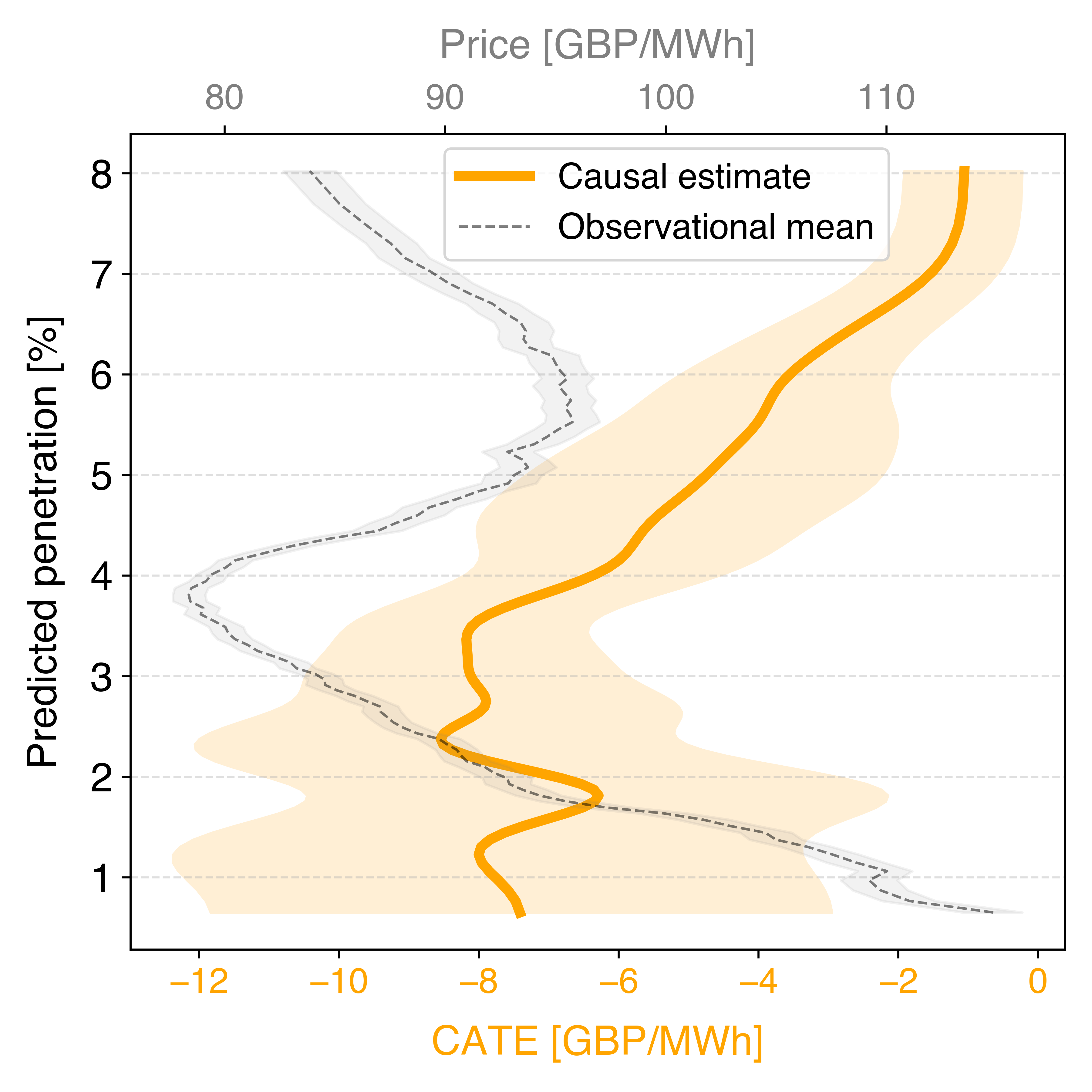}
        \caption{}
        \label{fig:dml-nordpool-b}
    \end{subfigure}
    \caption[Smoothed nonlinear effect for NordPool prices]{\textbf{Results from the NordPool day-ahead market}. Comparison of observational mean and causal effects of wind (a) and solar (b) power production on wholesale electricity prices. The solid lines represent the nonlinear CATE estimates derived using our locally partially linear DML framework, capturing the price impact [GBP/MWh] of a 1 GWh increase in renewable energy generation. The dashed lines show the observational mean trends as a function of renewable penetration levels, illustrating the differences between raw associations and the true causal effects. Shaded areas denote 80\% confidence intervals obtained with bootstrap procedures.}
    \label{fig:dml-nordpool}
\end{figure}

\paragraph{NordPool prices.} To establish the robustness of our locally partially linear DML framework, we extend the results presented in the paper to the NordPool (day-ahead) and the intraday prices. Here, we estimate the causal effects of wind and solar generation on electricity prices, highlighting differences between observed trends and true causal impacts. Figure \ref{fig:dml-nordpool} compares observational mean trends with smoothed nonlinear CATEs for wind and solar power on electricity prices. These results confirm the behaviour observed in the case of APX prices, where both wind and solar power generation exert a significant downward pressure on the electricity prices.

\begin{figure}[!ht]
    \centering
    \begin{subfigure}[b]{0.48\linewidth}
        \centering
        \includegraphics[width=\linewidth]{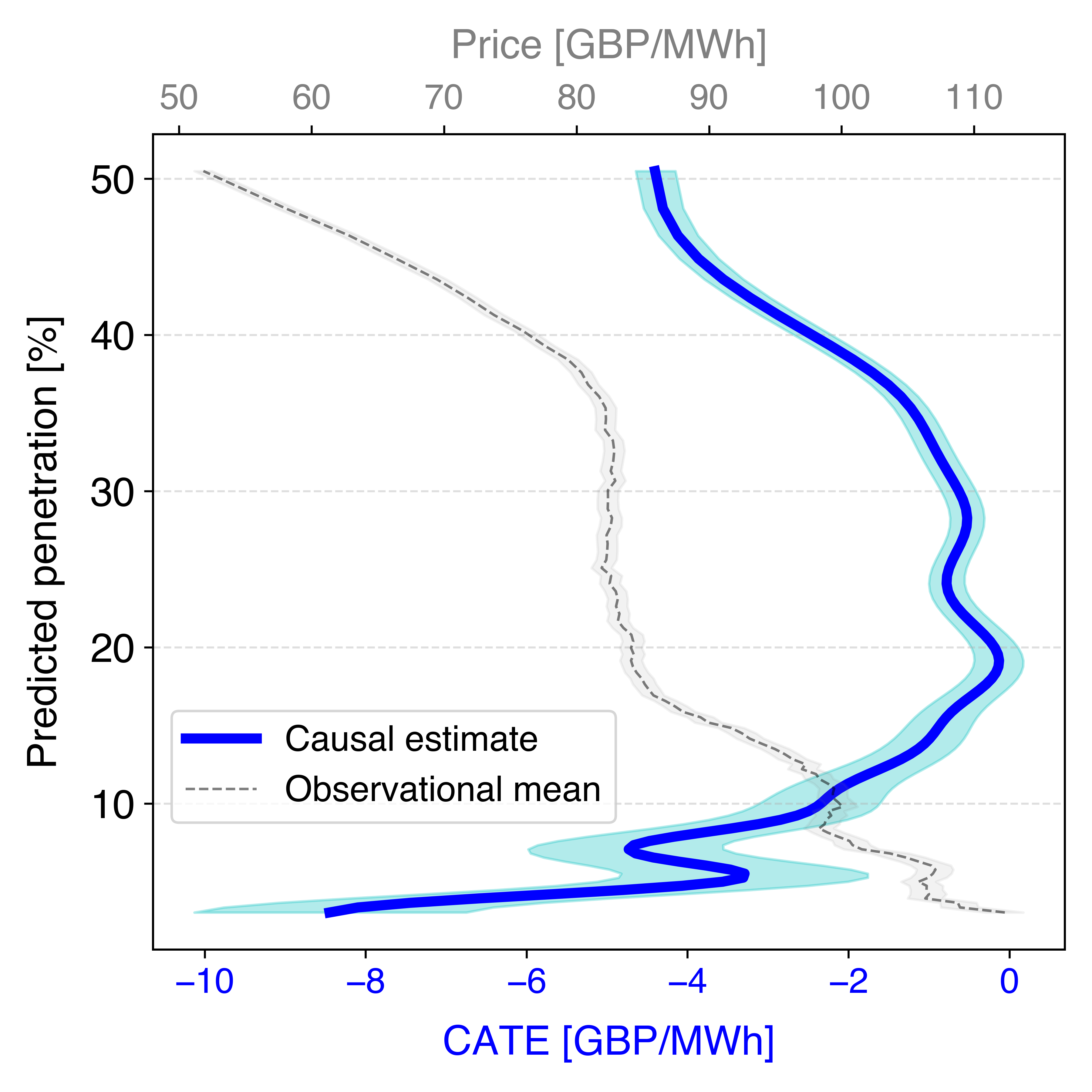}
        \caption{}
        \label{fig:dml-intraday-a}
    \end{subfigure}
    \hfill
    \begin{subfigure}[b]{0.48\linewidth}
        \centering
        \includegraphics[width=\linewidth]{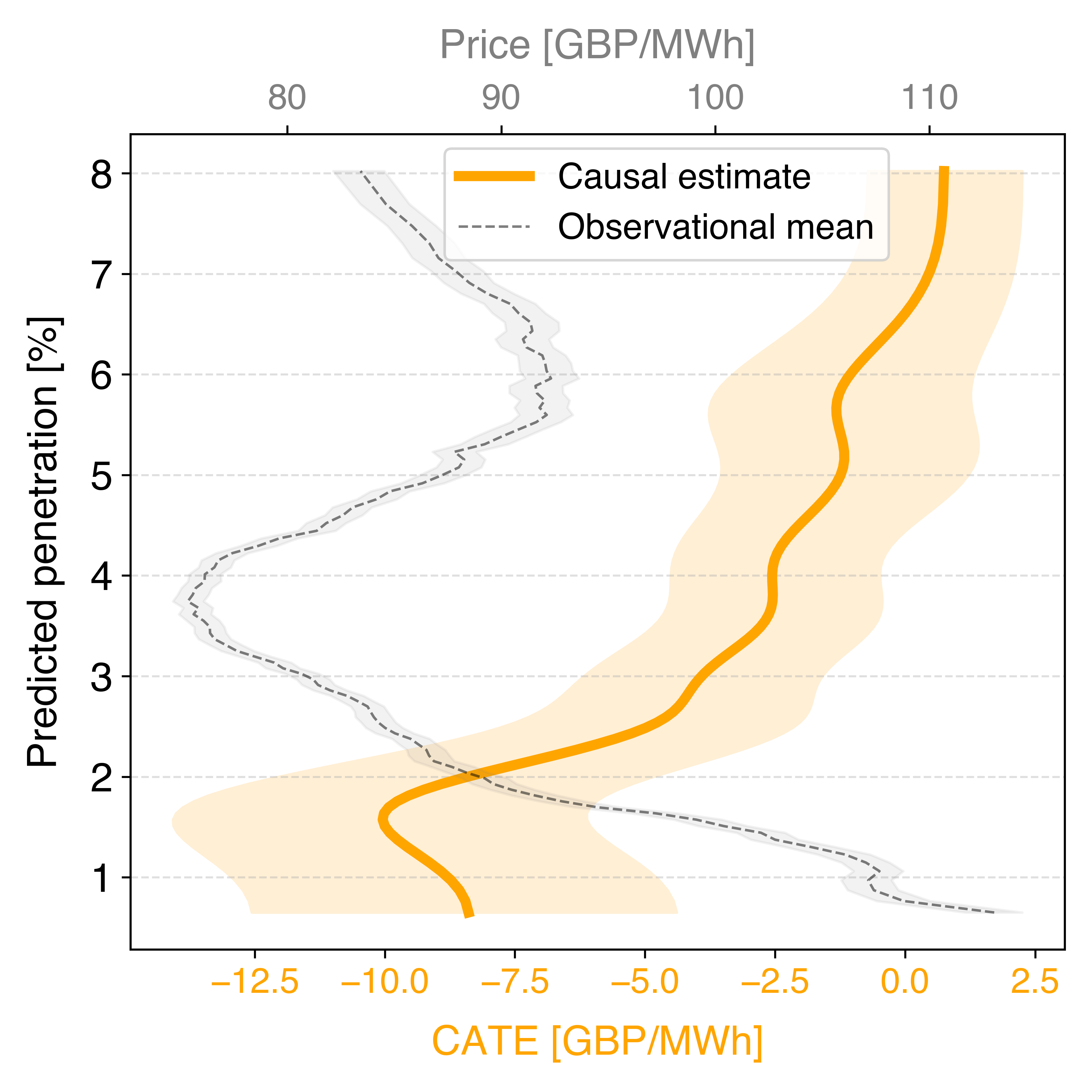}
        \caption{}
        \label{fig:dml-intraday-b}
    \end{subfigure}
    \caption[Smoothed nonlinear effect for intraday prices]{\textbf{Causal impact of predicted renewable power production on electricity prices (intraday market).} Comparison of observational mean and causal effects of wind (a) and solar (b) power production on wholesale electricity prices. The solid lines represent the nonlinear CATE estimates derived using our locally partially linear DML framework, capturing the price impact (GBP/MWh) of a 1 GWh increase in renewable energy generation. The dashed lines show the observational mean trends as a function of renewable penetration levels, illustrating the differences between raw associations and the true causal effects. Shaded areas denote 80\% confidence intervals obtained with bootstrap procedures.}
    \label{fig:dml-intraday}
\end{figure}

\clearpage
\paragraph{Intraday prices.} We further extend the analysis to the intraday market, as shown in Figure \ref{fig:dml-intraday}, where the smoothed CATE estimates illustrate the impacts of predicted renewable power generation on intraday electricity prices. Here, it is possible to notice how while wind power seems to have a quite comparable behaviour, solar power seems to have a much reduced effect on intraday prices, compared to the day-ahead markets.

\clearpage
\subsubsection*{Using penetration as input}
\begin{figure}[!ht]
    \centering
    \begin{subfigure}[b]{0.48\linewidth}
        \centering
        \includegraphics[width=\linewidth]{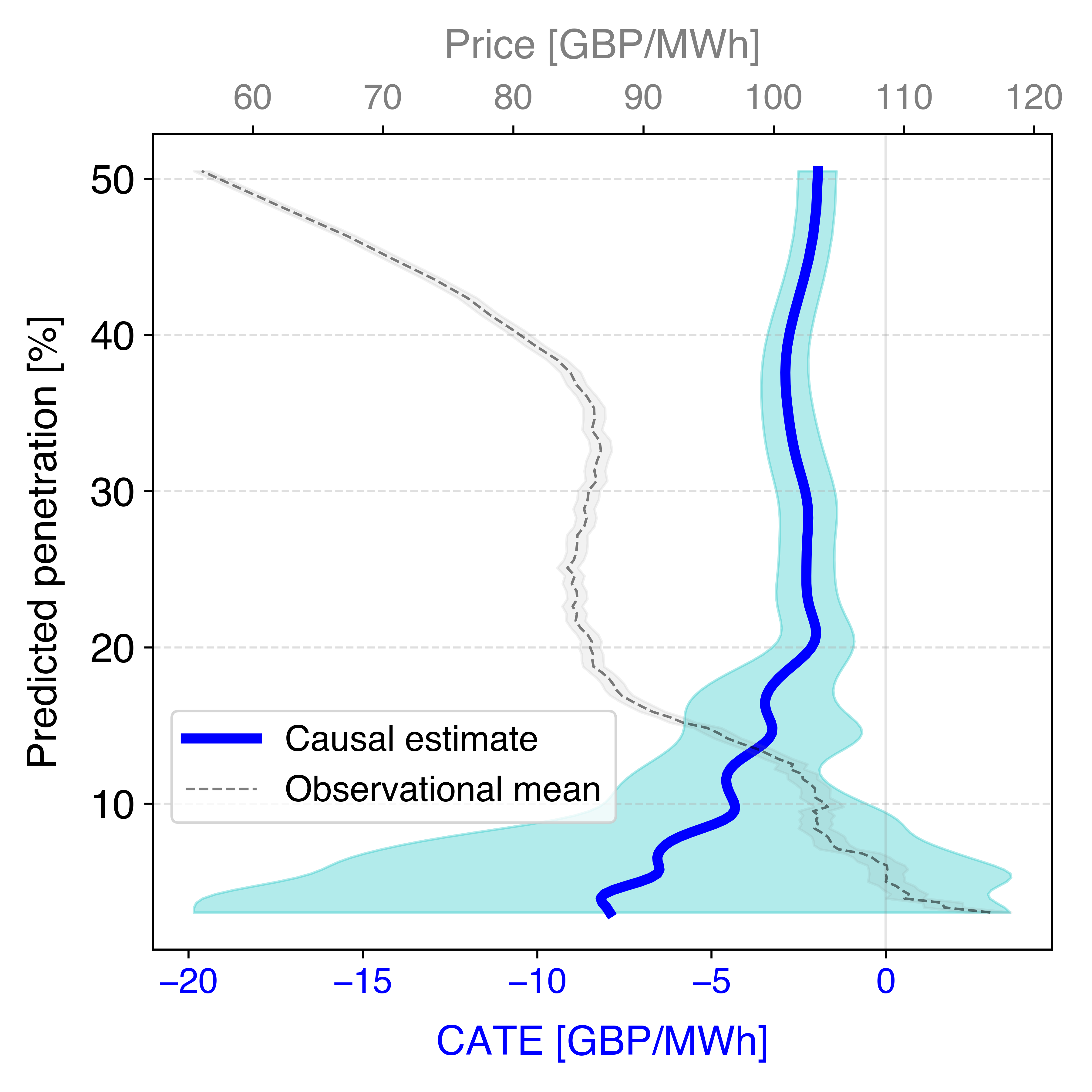}
        \caption{}
        \label{fig:dml-v3-a}
    \end{subfigure}
    \hfill
    \begin{subfigure}[b]{0.48\linewidth}
        \centering
        \includegraphics[width=\linewidth]{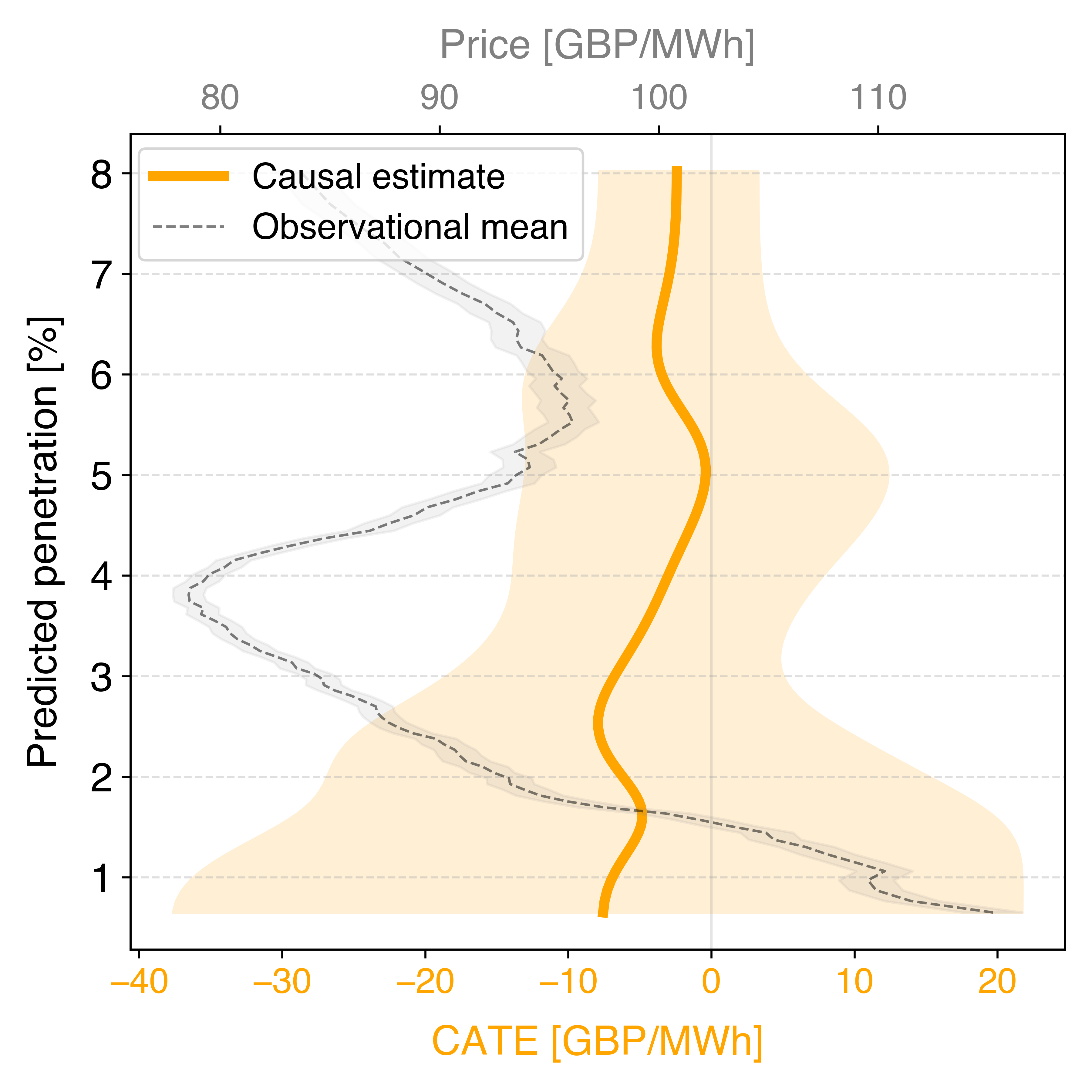}
        \caption{}
        \label{fig:dml-v3-b}
    \end{subfigure}
    \caption[Smoothed nonlinear effect of the predicted penetration]{\textbf{Effect of the predicted penetrations}. Nonlinear CATE estimates derived using our locally partially linear DML framework, capturing the price impact [GBP/MWh] of a 1\% increase in renewable penetration.}
    \label{fig:dml-v3}
\end{figure}

Here, we attempt to estimate the CATE using predicted renewable penetration as the input variable. Previously, predicted penetration was used as a contextual variable within the boxcar kernel, slicing the data into subsets to estimate localised effects. In that framework, the CATE represented the price impact of adding 1 extra GWh of renewable energy generation to the system. Here, we present an alternative analysis where the CATE reflects the price impact of increasing renewable penetration by 1 percentage point (\%). Figure \ref{fig:dml-v3} shows the results of applying our DML framework with predicted renewable penetration serving both as the input variable for the final model and as a contextual variable for the boxcar kernel. This approach introduces potential challenges, particularly regarding additional confounding factors. Normalising the predicted renewable production by the total estimated load to compute predicted penetration may introduce dependencies that are not fully accounted for by the standard confounders used in residualising production. This added confounding underscores the need for caution when interpreting these results. Unlike penetration, renewable production forecasts are primarily driven by well-understood factors such as weather conditions and seasonal patterns, making them less susceptible to hidden confounders. To mitigate these issues, we applied residualisation to predicted penetration using the same confounders employed for renewable production—namely, wind (or solar) capacity, daylight hours, hour of the day, and month of the year. Additionally, we included estimated load and predicted renewable production (wind or solar) as confounders to address dependencies introduced by the penetration calculation.

In Figure \ref{fig:dml-v3}, the smoothing factor of the Gaussian filter was set equal to the standard deviation of the mean CATE. For both wind and solar, we observe increased variability in the estimates. For wind, this variability is particularly pronounced below 10\% penetration. For solar, the variability remains consistently higher across the entire range of predicted penetration. This heightened variability is likely due to the smaller effect size, making the estimates more susceptible to confounding factors introduced during the normalisation process. Nevertheless, the mean causal estimates remain negative throughout the range, reaffirming the merit order effect, whereby increased renewable penetration reduces electricity prices.

\clearpage
\subsubsection*{Individual effect estimates prior to the smoothing}
\begin{figure}[!ht]
    \centering
    \begin{subfigure}[b]{0.48\linewidth}
        \centering
        \includegraphics[width=\linewidth]{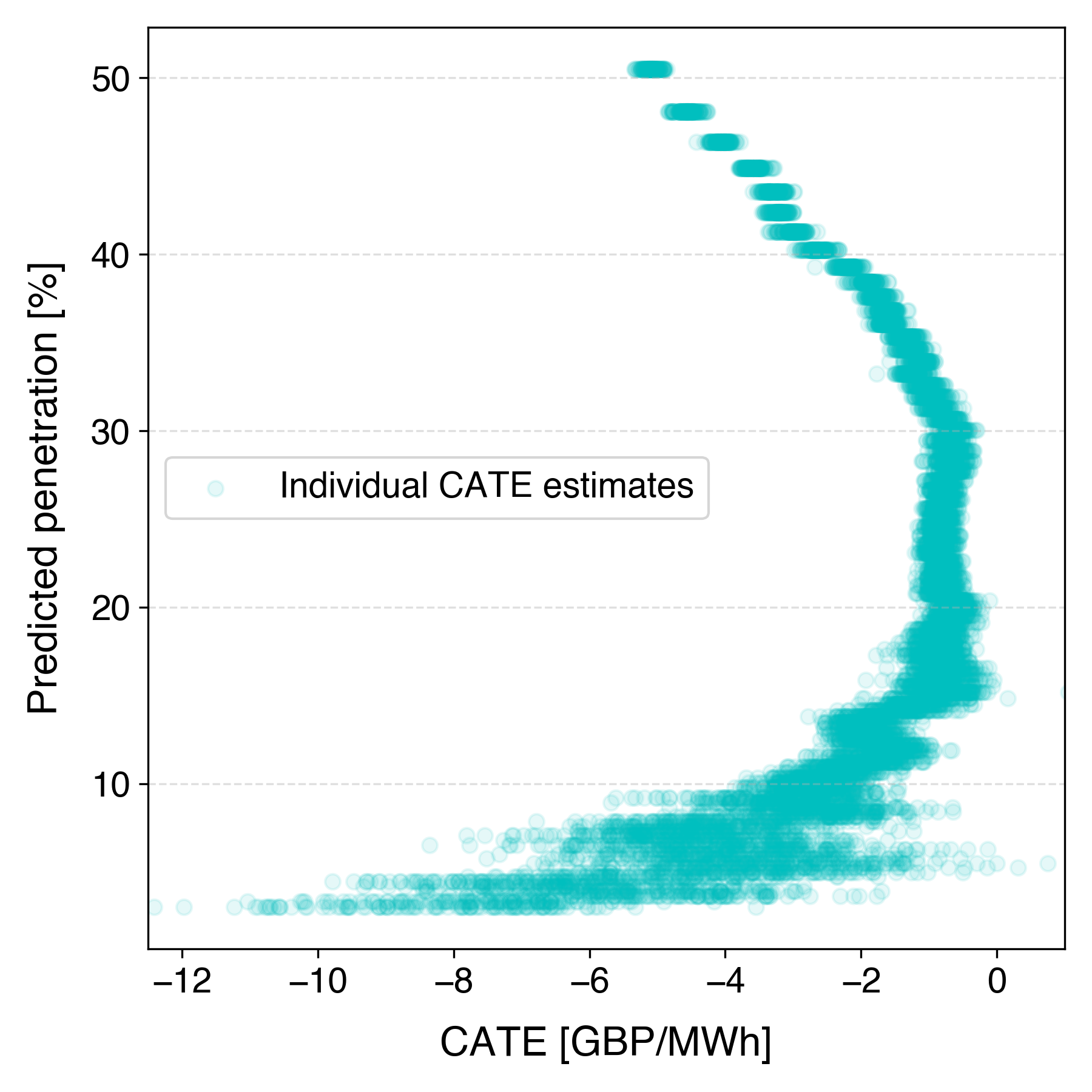}
        \caption{}
        \label{fig:dml-scatter-a}
    \end{subfigure}
    \hfill
    \begin{subfigure}[b]{0.48\linewidth}
        \centering
        \includegraphics[width=\linewidth]{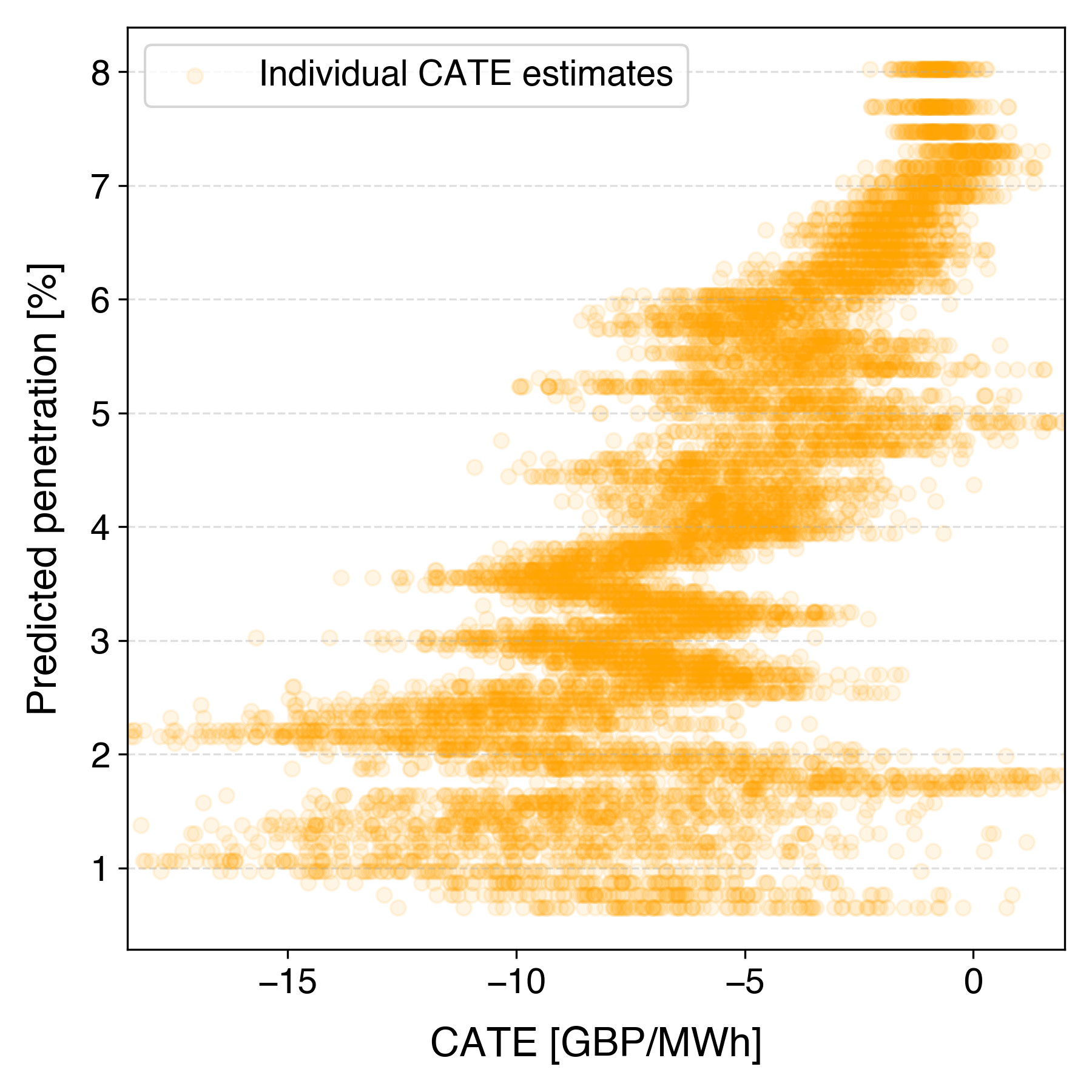}
        \caption{}
        \label{fig:dml-scatter-b}
    \end{subfigure}
    \caption[Scatterplot of individual CATE estimates for APX prices (overall analysis)]{\textbf{Individual CATE estimates prior to smoothing.} Individual CATE estimates of the effect of wind (a) and solar (b) power production on electricity prices.}
    \label{fig:dml-scatter}
\end{figure}

\begin{figure}[!ht]
    \centering
    \begin{subfigure}[b]{0.48\linewidth}
        \centering
        \includegraphics[width=\linewidth]{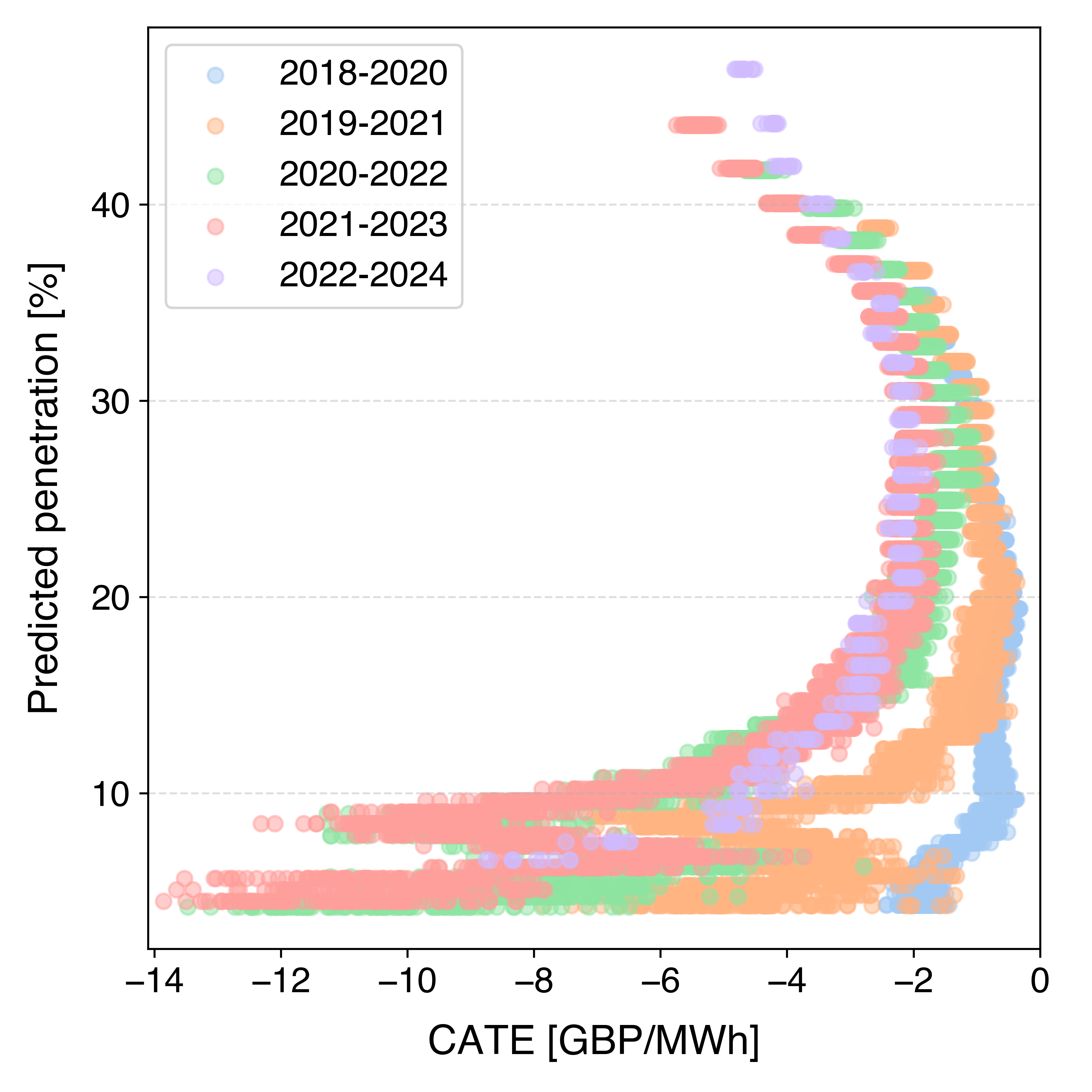}
        \caption{}
        \label{fig:dml-scatter-yearly-a}
    \end{subfigure}
    \hfill
    \begin{subfigure}[b]{0.48\linewidth}
        \centering
        \includegraphics[width=\linewidth]{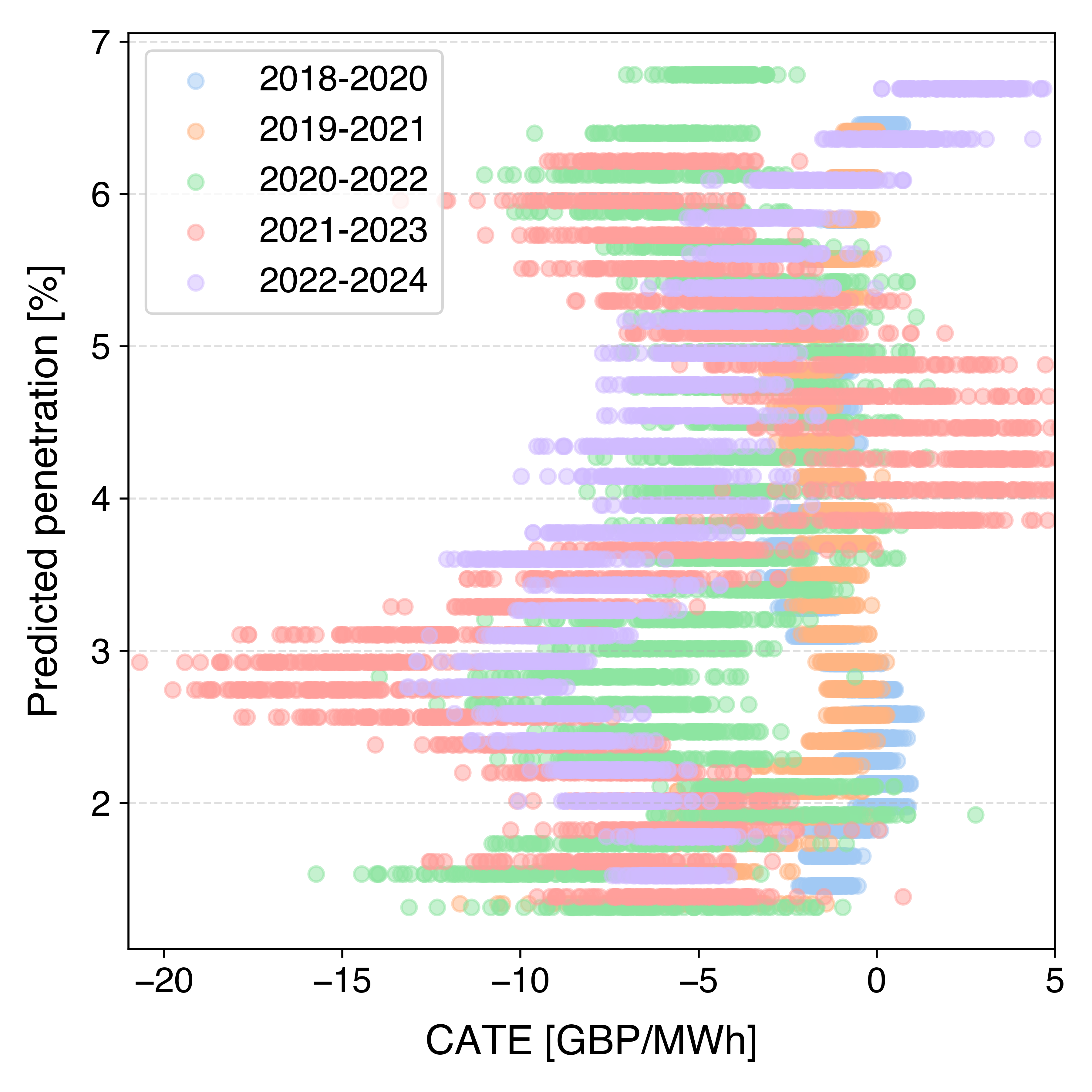}
        \caption{}
        \label{fig:dml-scatter-yearly-b}
    \end{subfigure}
    \caption[Scatterplot of individual CATE estimates for APX prices (yearly analysis)]{Temporal evolution of the individual CATE estimates of wind (a) and solar (b) power production on electricity prices.}
    \label{fig:dml-scatter-yearly}
\end{figure}

We now provide a visualisation for the individual CATE estimates obtained from each bootstrap iteration of our causal inference framework. These estimates represent the causal impact, in GBP/MWh, of a 1 GWh increase in renewable energy generation. Each data point corresponds to the output of our locally partially linear DML method prior to any smoothing. Figure \ref{fig:dml-scatter} illustrates the distribution of individual CATE estimates for wind and solar generation. Meanwhile, Figure \ref{fig:dml-scatter-yearly} shows the temporal evolution of these estimates across all windows, highlighting variability over time.

\clearpage
\subsection*{Supplementary tables}

Table \ref{tab:dataset_description} provides a comprehensive overview of the variables employed in our DML framework, emphasising their data sources and roles across different models. The dataset integrates observations from the EnAppSys platform\cite{enappsys}, direct inputs from the National Energy System Operator (NESO), and publicly available data from the NESO Data Portal\cite{nesodataportal}.

\begin{table}[!ht]
\centering
\small
\begin{tabular}{|m{2.2cm}|m{4.5cm}|m{1.5cm}|c|c|c|c|}
\hline
\textbf{Variable}                   & \textbf{Description}                                                & \textbf{Source} & \makecell{\textbf{Price} \\\textbf{(wind model)}} & \makecell{\textbf{Wind} \\ \textbf{forecast}} & \makecell{\textbf{Price} \\ \textbf{(solar model)}} & \makecell{\textbf{Solar} \\ \textbf{forecast}}\\ \hline
\texttt{Date}                        & Date of the observation (datetime index). One observation for each 30-minute settlement period. & EnAppSys & $\times$ & $\times$ & $\times$ & $\times$ \\ \hline
\texttt{Year}                        & Year of observation.                                                 & Derived & \checkmark & $\times$ & \checkmark & $\times$ \\ \hline
\texttt{Month}                       & Month of the year (1-12).                                             & Derived & \checkmark & \checkmark & \checkmark & \checkmark \\ \hline
\texttt{Day}                         & Day of the week (0-6, Monday to Sunday).                             & Derived & \checkmark & $\times$ & \checkmark & $\times$ \\ \hline
\texttt{Hour}                        & Hour of the day (0-23).                                              & Derived & \checkmark & \checkmark & \checkmark & \checkmark \\ \hline
\texttt{Daylight hours}              & Number of hours of daylight based on geographical location. It has been computed using the Python Astral package \cite{astral2023}. & Derived & \checkmark & \checkmark & \checkmark & \checkmark \\ \hline
\texttt{APX price}        & Day-ahead electricity price on the APX exchange [GBP/MWh].            & EnAppSys  & $\times$ & $\times$ & $\times$ & $\times$ \\ \hline
\texttt{NordPool price}             & Day-ahead electricity price on the NordPool exchange [GBP/MWh].        & EnAppSys     & $\times$ & $\times$ & $\times$ & $\times$ \\ \hline
\texttt{Intraday price}          & Intraday or within-day (MIDP) electricity price [GBP/MWh]. & EnAppSys     & $\times$ & $\times$ & $\times$ & $\times$ \\ \hline
\texttt{Actual load}             & Initial transmission system demand outturn [MW].  & EnAppSys & $\times$ & $\times$ & $\times$ & $\times$ \\ \hline
\texttt{Estimated load}             & Estimated electricity load, generated from actual demand with noise [MW]. & Derived & \checkmark & $\times$ & \checkmark & $\times$ \\ \hline
\texttt{Gas price}                   & National Balancing Point (NBP) price for natural gas [GBP/MWh].               & EnAppSys & \checkmark & $\times$ & \checkmark & $\times$ \\ \hline
\texttt{Carbon permits}                   & MID CO\textsubscript{2} prices from the EU Emissions Trading Scheme [GBP/tCO\textsubscript{2}e]. & EnAppSys & \checkmark & $\times$ & \checkmark & $\times$ \\ \hline
\texttt{Wind capacity}              & Installed wind capacity [MW].                                    & NESO & $\times$ & \checkmark & $\times$ & $\times$ \\ \hline
\texttt{Wind forecast}              & Predicted wind production [MW].                                     & NESO & $\times$ & $\times$ & \checkmark & $\times$ \\ \hline
\texttt{Predicted wind penetration} & Obtained as: wind forecast / estimated load * 100 [\%].    & Derived & $\times$ & $\times$ & $\times$ & $\times$ \\ \hline
\texttt{Solar capacity}             & Installed solar capacity [MW].                                   & NESO & $\times$ & $\times$ & $\times$ & \checkmark \\ \hline
\texttt{Solar forecast}             & Predicted solar production [MW].                                    & NESO & \checkmark & $\times$ & $\times$ & $\times$ \\ \hline
\texttt{Predicted solar penetration} & Obtained as: solar forecast / estimated load * 100) [\%].  & Derived & $\times$ & $\times$ & $\times$ & $\times$ \\ \hline
\end{tabular}
\caption[Description of variables used in the machine learning models]{Dataset Description with variables used in the DML framework. The check marks (\checkmark) indicate whether the variable in the row was used to residualise the variable in the column. For example, if a variable is marked with a \checkmark\ under ``Price (wind model)'', it means this row's variable was used as a confounder in residualising the electricity price in the wind model.}
\label{tab:dataset_description}
\end{table}

\clearpage
\bibliography{ref}

\begin{thebibliography}{10}
\urlstyle{rm}
\expandafter\ifx\csname url\endcsname\relax
  \def\url#1{\texttt{#1}}\fi
\expandafter\ifx\csname urlprefix\endcsname\relax\def\urlprefix{URL }\fi
\expandafter\ifx\csname doiprefix\endcsname\relax\def\doiprefix{DOI: }\fi
\providecommand{\bibinfo}[2]{#2}
\providecommand{\eprint}[2][]{\url{#2}}

\bibitem{IEA2023}
\bibinfo{author}{IEA}.
\newblock \bibinfo{title}{Net zero roadmap: A global pathway to keep the 1.5 °c goal in reach -- 2023 update}.
\newblock \bibinfo{type}{Technical Report}, \bibinfo{institution}{International Energy Agency} (\bibinfo{year}{2023}).

\bibitem{nationalgrid2023_wind}
\bibinfo{author}{{National Grid}}.
\newblock \bibinfo{title}{How much of the uk's energy is renewable} (\bibinfo{year}{2023}).
\newblock \bibinfo{note}{Accessed: 2023-04-12}.

\bibitem{IPCC2023Summary}
\bibinfo{author}{{IPCC}}.
\newblock \bibinfo{title}{Climate change synthesis report 2023 -- summary for policymakers}.
\newblock In \bibinfo{editor}{Lee, H.} \& \bibinfo{editor}{Romero, J.} (eds.) \emph{\bibinfo{booktitle}{Climate Change 2023: Synthesis Report. Contribution of Working Groups I, II and III to the Sixth Assessment Report of the Intergovernmental Panel on Climate Change}}, \bibinfo{pages}{1--34} (\bibinfo{publisher}{IPCC}, \bibinfo{address}{Geneva, Switzerland}, \bibinfo{year}{2023}).

\bibitem{morales2013integrating}
\bibinfo{author}{Morales, J.~M.}, \bibinfo{author}{Conejo, A.~J.}, \bibinfo{author}{Madsen, H.}, \bibinfo{author}{Pinson, P.} \& \bibinfo{author}{Zugno, M.}
\newblock \emph{\bibinfo{title}{Integrating renewables in electricity markets: operational problems}}, vol. \bibinfo{volume}{205} (\bibinfo{publisher}{Springer Science \& Business Media}, \bibinfo{year}{2013}).

\bibitem{antweiler2021long}
\bibinfo{author}{Antweiler, W.} \& \bibinfo{author}{Muesgens, F.}
\newblock \bibinfo{journal}{\bibinfo{title}{On the long-term merit order effect of renewable energies}}.
\newblock {\emph{\JournalTitle{Energy Economics}}} \textbf{\bibinfo{volume}{99}}, \bibinfo{pages}{105275} (\bibinfo{year}{2021}).

\bibitem{miettinen2019impacts}
\bibinfo{author}{Miettinen, J.} \& \bibinfo{author}{Holttinen, H.}
\newblock \bibinfo{journal}{\bibinfo{title}{Impacts of wind power forecast errors on the real-time balancing need: a {N}ordic case study}}.
\newblock {\emph{\JournalTitle{IET Renewable Power Generation}}} \textbf{\bibinfo{volume}{13}}, \bibinfo{pages}{227--233} (\bibinfo{year}{2019}).

\bibitem{goodarzi2019impact}
\bibinfo{author}{Goodarzi, S.}, \bibinfo{author}{Perera, H.~N.} \& \bibinfo{author}{Bunn, D.}
\newblock \bibinfo{journal}{\bibinfo{title}{The impact of renewable energy forecast errors on imbalance volumes and electricity spot prices}}.
\newblock {\emph{\JournalTitle{Energy Policy}}} \textbf{\bibinfo{volume}{134}}, \bibinfo{pages}{110827} (\bibinfo{year}{2019}).

\bibitem{morthorst2003wind}
\bibinfo{author}{Morthorst, P.~E.}
\newblock \bibinfo{journal}{\bibinfo{title}{Wind power and the conditions at a liberalized power market}}.
\newblock {\emph{\JournalTitle{Wind Energy}}} \textbf{\bibinfo{volume}{6}}, \bibinfo{pages}{297--308} (\bibinfo{year}{2003}).

\bibitem{jonsson2010market}
\bibinfo{author}{J{\'o}nsson, T.}, \bibinfo{author}{Pinson, P.} \& \bibinfo{author}{Madsen, H.}
\newblock \bibinfo{journal}{\bibinfo{title}{On the market impact of wind energy forecasts}}.
\newblock {\emph{\JournalTitle{Energy Economics}}} \textbf{\bibinfo{volume}{32}}, \bibinfo{pages}{313--320} (\bibinfo{year}{2010}).

\bibitem{cutler2011high}
\bibinfo{author}{Cutler, N.~J.}, \bibinfo{author}{Boerema, N.~D.}, \bibinfo{author}{MacGill, I.~F.} \& \bibinfo{author}{Outhred, H.~R.}
\newblock \bibinfo{journal}{\bibinfo{title}{High penetration wind generation impacts on spot prices in the {A}ustralian national electricity market}}.
\newblock {\emph{\JournalTitle{Energy Policy}}} \textbf{\bibinfo{volume}{39}}, \bibinfo{pages}{5939--5949} (\bibinfo{year}{2011}).

\bibitem{forrest2013assessing}
\bibinfo{author}{Forrest, S.} \& \bibinfo{author}{MacGill, I.}
\newblock \bibinfo{journal}{\bibinfo{title}{Assessing the impact of wind generation on wholesale prices and generator dispatch in the {A}ustralian {N}ational {E}lectricity {M}arket}}.
\newblock {\emph{\JournalTitle{Energy Policy}}} \textbf{\bibinfo{volume}{59}}, \bibinfo{pages}{120--132} (\bibinfo{year}{2013}).

\bibitem{bell2017revitalising}
\bibinfo{author}{Bell, W.~P.}, \bibinfo{author}{Wild, P.}, \bibinfo{author}{Foster, J.} \& \bibinfo{author}{Hewson, M.}
\newblock \bibinfo{journal}{\bibinfo{title}{Revitalising the wind power induced merit order effect to reduce wholesale and retail electricity prices in {A}ustralia}}.
\newblock {\emph{\JournalTitle{Energy Economics}}} \textbf{\bibinfo{volume}{67}}, \bibinfo{pages}{224--241} (\bibinfo{year}{2017}).

\bibitem{wurzburg2013renewable}
\bibinfo{author}{W{\"u}rzburg, K.}, \bibinfo{author}{Labandeira, X.} \& \bibinfo{author}{Linares, P.}
\newblock \bibinfo{journal}{\bibinfo{title}{Renewable generation and electricity prices: Taking stock and new evidence for {G}ermany and {A}ustria}}.
\newblock {\emph{\JournalTitle{Energy Economics}}} \textbf{\bibinfo{volume}{40}}, \bibinfo{pages}{S159--S171} (\bibinfo{year}{2013}).

\bibitem{ketterer2014impact}
\bibinfo{author}{Ketterer, J.~C.}
\newblock \bibinfo{journal}{\bibinfo{title}{The impact of wind power generation on the electricity price in {G}ermany}}.
\newblock {\emph{\JournalTitle{Energy economics}}} \textbf{\bibinfo{volume}{44}}, \bibinfo{pages}{270--280} (\bibinfo{year}{2014}).

\bibitem{gurtler2018effect}
\bibinfo{author}{G{\"u}rtler, M.} \& \bibinfo{author}{Paulsen, T.}
\newblock \bibinfo{journal}{\bibinfo{title}{The effect of wind and solar power forecasts on day-ahead and intraday electricity prices in {G}ermany}}.
\newblock {\emph{\JournalTitle{Energy Economics}}} \textbf{\bibinfo{volume}{75}}, \bibinfo{pages}{150--162} (\bibinfo{year}{2018}).

\bibitem{do2019impact}
\bibinfo{author}{Do, L. P.~C.}, \bibinfo{author}{Ly{\'o}csa, {\v{S}}.} \& \bibinfo{author}{Moln{\'a}r, P.}
\newblock \bibinfo{journal}{\bibinfo{title}{Impact of wind and solar production on electricity prices: Quantile regression approach}}.
\newblock {\emph{\JournalTitle{Journal of the Operational Research Society}}} \textbf{\bibinfo{volume}{70}}, \bibinfo{pages}{1752--1768} (\bibinfo{year}{2019}).

\bibitem{keeley2021impact}
\bibinfo{author}{Keeley, A.~R.}, \bibinfo{author}{Matsumoto, K.}, \bibinfo{author}{Tanaka, K.}, \bibinfo{author}{Sugiawan, Y.} \& \bibinfo{author}{Managi, S.}
\newblock \bibinfo{journal}{\bibinfo{title}{The impact of renewable energy generation on the spot market price in {G}ermany: ex-post analysis using boosting method}}.
\newblock {\emph{\JournalTitle{The Energy Journal}}} \textbf{\bibinfo{volume}{42}}, \bibinfo{pages}{1--22} (\bibinfo{year}{2021}).

\bibitem{o2011merit}
\bibinfo{author}{O'Mahoney, A.} \& \bibinfo{author}{Denny, E.}
\newblock \bibinfo{journal}{\bibinfo{title}{The merit order effect of wind generation on the irish electricity market}}.
\newblock {\emph{\JournalTitle{MPRA Paper}}}  (\bibinfo{year}{2011}).

\bibitem{o2014quantitative}
\bibinfo{author}{O’Flaherty, M.}, \bibinfo{author}{Riordan, N.}, \bibinfo{author}{O’Neill, N.} \& \bibinfo{author}{Ahern, C.}
\newblock \bibinfo{journal}{\bibinfo{title}{A quantitative analysis of the impact of wind energy penetration on electricity prices in {I}reland}}.
\newblock {\emph{\JournalTitle{Energy Procedia}}} \textbf{\bibinfo{volume}{58}}, \bibinfo{pages}{103--110} (\bibinfo{year}{2014}).

\bibitem{clo2015merit}
\bibinfo{author}{Cl{\`o}, S.}, \bibinfo{author}{Cataldi, A.} \& \bibinfo{author}{Zoppoli, P.}
\newblock \bibinfo{journal}{\bibinfo{title}{The merit-order effect in the {I}talian power market: The impact of solar and wind generation on national wholesale electricity prices}}.
\newblock {\emph{\JournalTitle{Energy Policy}}} \textbf{\bibinfo{volume}{77}}, \bibinfo{pages}{79--88} (\bibinfo{year}{2015}).

\bibitem{shcherbakova2014effect}
\bibinfo{author}{Shcherbakova, A.}, \bibinfo{author}{Kleit, A.}, \bibinfo{author}{Blumsack, S.}, \bibinfo{author}{Cho, J.} \& \bibinfo{author}{Lee, W.}
\newblock \bibinfo{journal}{\bibinfo{title}{Effect of increased wind penetration on system prices in {K}orea's electricity markets}}.
\newblock {\emph{\JournalTitle{Wind Energy}}} \textbf{\bibinfo{volume}{17}}, \bibinfo{pages}{1469--1482} (\bibinfo{year}{2014}).

\bibitem{nieuwenhout2011impact}
\bibinfo{author}{Nieuwenhout, F.} \& \bibinfo{author}{Brand, A.}
\newblock \bibinfo{title}{The impact of wind power on day-ahead electricity prices in the {N}etherlands}.
\newblock In \emph{\bibinfo{booktitle}{2011 8th International Conference on the European Energy Market (EEM)}}, \bibinfo{pages}{226--230} (\bibinfo{organization}{IEEE}, \bibinfo{year}{2011}).

\bibitem{hu2021effects}
\bibinfo{author}{Hu, X.}, \bibinfo{author}{Jarait{\.e}, J.} \& \bibinfo{author}{Ka{\v{z}}ukauskas, A.}
\newblock \bibinfo{journal}{\bibinfo{title}{The effects of wind power on electricity markets: A case study of the {S}wedish intraday market}}.
\newblock {\emph{\JournalTitle{Energy Economics}}} \textbf{\bibinfo{volume}{96}}, \bibinfo{pages}{105159} (\bibinfo{year}{2021}).

\bibitem{spodniak2021impact}
\bibinfo{author}{Spodniak, P.}, \bibinfo{author}{Ollikka, K.} \& \bibinfo{author}{Honkapuro, S.}
\newblock \bibinfo{journal}{\bibinfo{title}{The impact of wind power and electricity demand on the relevance of different short-term electricity markets: The {N}ordic case}}.
\newblock {\emph{\JournalTitle{Applied energy}}} \textbf{\bibinfo{volume}{283}}, \bibinfo{pages}{116063} (\bibinfo{year}{2021}).

\bibitem{tselika2024quantifying}
\bibinfo{author}{Tselika, K.}, \bibinfo{author}{Tselika, M.} \& \bibinfo{author}{Demetriades, E.}
\newblock \bibinfo{journal}{\bibinfo{title}{Quantifying the short-term asymmetric effects of renewable energy on the electricity merit-order curve}}.
\newblock {\emph{\JournalTitle{Energy Economics}}} \textbf{\bibinfo{volume}{132}}, \bibinfo{pages}{107471} (\bibinfo{year}{2024}).

\bibitem{gelabert2011ex}
\bibinfo{author}{Gelabert, L.}, \bibinfo{author}{Labandeira, X.} \& \bibinfo{author}{Linares, P.}
\newblock \bibinfo{journal}{\bibinfo{title}{An ex-post analysis of the effect of renewables and cogeneration on {S}panish electricity prices}}.
\newblock {\emph{\JournalTitle{Energy economics}}} \textbf{\bibinfo{volume}{33}}, \bibinfo{pages}{S59--S65} (\bibinfo{year}{2011}).

\bibitem{cruz2011effect}
\bibinfo{author}{Cruz, A.}, \bibinfo{author}{Mu{\~n}oz, A.}, \bibinfo{author}{Zamora, J.~L.} \& \bibinfo{author}{Esp{\'\i}nola, R.}
\newblock \bibinfo{journal}{\bibinfo{title}{The effect of wind generation and weekday on {S}panish electricity spot price forecasting}}.
\newblock {\emph{\JournalTitle{Electric Power Systems Research}}} \textbf{\bibinfo{volume}{81}}, \bibinfo{pages}{1924--1935} (\bibinfo{year}{2011}).

\bibitem{azofra2014wind}
\bibinfo{author}{Azofra, D.}, \bibinfo{author}{Jim{\'e}nez, E.}, \bibinfo{author}{Mart{\'\i}nez, E.}, \bibinfo{author}{Blanco, J.} \& \bibinfo{author}{Saenz-D{\'\i}ez, J.}
\newblock \bibinfo{journal}{\bibinfo{title}{Wind power merit-order and feed-in-tariffs effect: A variability analysis of the {S}panish electricity market}}.
\newblock {\emph{\JournalTitle{Energy Conversion and Management}}} \textbf{\bibinfo{volume}{83}}, \bibinfo{pages}{19--27} (\bibinfo{year}{2014}).

\bibitem{barthelmie2008economic}
\bibinfo{author}{Barthelmie, R.}, \bibinfo{author}{Murray, F.} \& \bibinfo{author}{Pryor, S.}
\newblock \bibinfo{journal}{\bibinfo{title}{The economic benefit of short-term forecasting for wind energy in the {UK} electricity market}}.
\newblock {\emph{\JournalTitle{Energy Policy}}} \textbf{\bibinfo{volume}{36}}, \bibinfo{pages}{1687--1696} (\bibinfo{year}{2008}).

\bibitem{joos2018short}
\bibinfo{author}{Joos, M.} \& \bibinfo{author}{Staffell, I.}
\newblock \bibinfo{journal}{\bibinfo{title}{Short-term integration costs of variable renewable energy: Wind curtailment and balancing in {B}ritain and {G}ermany}}.
\newblock {\emph{\JournalTitle{Renewable and Sustainable Energy Reviews}}} \textbf{\bibinfo{volume}{86}}, \bibinfo{pages}{45--65} (\bibinfo{year}{2018}).

\bibitem{woo2011impact}
\bibinfo{author}{Woo, C.-K.}, \bibinfo{author}{Horowitz, I.}, \bibinfo{author}{Moore, J.} \& \bibinfo{author}{Pacheco, A.}
\newblock \bibinfo{journal}{\bibinfo{title}{The impact of wind generation on the electricity spot-market price level and variance: {T}he {T}exas experience}}.
\newblock {\emph{\JournalTitle{Energy Policy}}} \textbf{\bibinfo{volume}{39}}, \bibinfo{pages}{3939--3944} (\bibinfo{year}{2011}).

\bibitem{woo2016merit}
\bibinfo{author}{Woo, C.-K.} \emph{et~al.}
\newblock \bibinfo{journal}{\bibinfo{title}{Merit-order effects of renewable energy and price divergence in {C}alifornia’s day-ahead and real-time electricity markets}}.
\newblock {\emph{\JournalTitle{Energy Policy}}} \textbf{\bibinfo{volume}{92}}, \bibinfo{pages}{299--312} (\bibinfo{year}{2016}).

\bibitem{martinez2016impact}
\bibinfo{author}{Martinez-Anido, C.~B.}, \bibinfo{author}{Brinkman, G.} \& \bibinfo{author}{Hodge, B.-M.}
\newblock \bibinfo{journal}{\bibinfo{title}{The impact of wind power on electricity prices}}.
\newblock {\emph{\JournalTitle{Renewable energy}}} \textbf{\bibinfo{volume}{94}}, \bibinfo{pages}{474--487} (\bibinfo{year}{2016}).

\bibitem{prol2020cannibalization}
\bibinfo{author}{Prol, J.~L.}, \bibinfo{author}{Steininger, K.~W.} \& \bibinfo{author}{Zilberman, D.}
\newblock \bibinfo{journal}{\bibinfo{title}{The cannibalization effect of wind and solar in the {C}alifornia wholesale electricity market}}.
\newblock {\emph{\JournalTitle{Energy Economics}}} \textbf{\bibinfo{volume}{85}}, \bibinfo{pages}{104552} (\bibinfo{year}{2020}).

\bibitem{stiewe2024cross}
\bibinfo{author}{Stiewe, C.}, \bibinfo{author}{Xu, A.~L.}, \bibinfo{author}{Eicke, A.} \& \bibinfo{author}{Hirth, L.}
\newblock \bibinfo{journal}{\bibinfo{title}{Cross-border cannibalization: Spillover effects of wind and solar energy on interconnected {E}uropean electricity markets}}.
\newblock {\emph{\JournalTitle{arXiv preprint arXiv:2405.17166}}}  (\bibinfo{year}{2024}).

\bibitem{chernozhukov2018double}
\bibinfo{author}{Chernozhukov, V.} \emph{et~al.}
\newblock \bibinfo{title}{Double/debiased machine learning for treatment and structural parameters} (\bibinfo{year}{2018}).

\bibitem{pearl2009causality}
\bibinfo{author}{Pearl, J.}
\newblock \emph{\bibinfo{title}{Causality}} (\bibinfo{publisher}{Cambridge university press}, \bibinfo{year}{2009}).

\bibitem{peters2017}
\bibinfo{author}{Peters, J.}, \bibinfo{author}{Janzing, D.} \& \bibinfo{author}{Sch{\"o}lkopf, B.}
\newblock \emph{\bibinfo{title}{Elements of Causal Inference: Foundations and Learning Algorithms}} (\bibinfo{publisher}{MIT Press}, \bibinfo{year}{2017}).

\bibitem{fuhr2024estimating}
\bibinfo{author}{Fuhr, J.}, \bibinfo{author}{Berens, P.} \& \bibinfo{author}{Papies, D.}
\newblock \bibinfo{journal}{\bibinfo{title}{Estimating causal effects with double machine learning--a method evaluation}}.
\newblock {\emph{\JournalTitle{arXiv preprint arXiv:2403.14385}}}  (\bibinfo{year}{2024}).

\bibitem{ke2017lightgbm}
\bibinfo{author}{Ke, G.} \emph{et~al.}
\newblock \bibinfo{title}{Light{GBM}: A highly efficient gradient boosting decision tree}.
\newblock In \emph{\bibinfo{booktitle}{Proceedings of the 31st International Conference on Neural Information Processing Systems}}, \bibinfo{pages}{3149--3157} (\bibinfo{year}{2017}).

\bibitem{enappsys}
\bibinfo{author}{{EnAppSys}}.
\newblock \bibinfo{title}{{EnAppSys} market data platform}.
\newblock \bibinfo{howpublished}{\url{https://www.enappsys.com/}} (\bibinfo{year}{2024}).

\bibitem{nesodataportal}
\bibinfo{title}{{NESO} data portal}.
\newblock \bibinfo{howpublished}{\url{https://www.neso.energy/data-portal}} (\bibinfo{year}{2024}).

\bibitem{astral2023}
\bibinfo{author}{Kennedy, S.}
\newblock \bibinfo{title}{Astral: a python package for calculating the times of various aspects of the sun and moon}.
\newblock \bibinfo{howpublished}{\url{https://sffjunkie.github.io/astral/}} (\bibinfo{year}{2023}).
\newblock \bibinfo{note}{Version 3.3.2, Python library for astronomical calculations}.

\end{thebibliography}

\end{document}